\documentclass[preprint,sort&compress]{elsarticle}
\usepackage{amssymb,amsthm,amsmath,commath,bbm,psfrag,setspace,color,array,booktabs,multirow,url,float}
\usepackage[FIGTOPCAP,nooneline]{subfigure}
 \usepackage{color}

 \usepackage{numcompress}

\definecolor{darkred}{rgb}{0.5,0,0}

\numberwithin{equation}{section}
\numberwithin{figure}{section}
\numberwithin{table}{section}

\theoremstyle{definition}
\newtheorem{definition}{Definition}[section]

\newtheorem{theorem}{Theorem}[section]

\newtheorem{lemma}[theorem]{Lemma}
\newtheorem{prop}{Proposition}
\newtheorem*{remark}{Remark}

\begin{document}

\begin{frontmatter}

\title{Calculating the Lyapunov exponents of a piecewise-smooth soft impacting system with a time-delayed feedback controller}

\author{Zhi Zhang}
\ead{zz326@exeter.ac.uk}

\author{Yang Liu$^\ast$}
\ead{y.liu2@exeter.ac.uk}

\author{Jan Sieber}
\ead{j.sieber@exeter.ac.uk}

\address[EXT]{College of Engineering Mathematics and
Physical Sciences, University of Exeter, Harrison Building, North
Park Road, Exeter EX4 4QF, UK}

\begin{abstract}
Lyapunov exponents are a widely used tool for studying dynamical systems. When calculating Lyapunov exponents for piecewise-smooth systems with time-delayed arguments one faces a lack of continuity in the variational problem. This paper studies how to build a variational equation for the efficient construction of Jacobians along trajectories of the delayed nonsmooth system. Trajectories of the piecewise-smooth system may encounter a so-called grazing event where the trajectory approaches a discontinuity surface in the state space in a non-transversal manner. For this event we develop a grazing point estimation algorithm to ensure the accuracy of trajectories for the nonlinear and the variational equations. We show that the eigenvalues of the Jacobian matrix computed by the algorithm converge with an order consistent with the order of the numerical integration method, therefore guaranteeing the reliability of our proposed numerical method. Finally, the method is demonstrated on a periodically forced impacting oscillator under the time-delayed feedback control.
\end{abstract}

\begin{keyword}
Lyapunov exponents; Piecewise-smooth dynamical system; Delay differential equation; Grazing; Impact oscillator.
\end{keyword}

\end{frontmatter}

\section{\label{sec_intro}Introduction}

Analysing grazing events for nonsmooth systems is a challenging task \cite{bernardo2008piecewise}.
In general, vibro-impact systems, such as ship mooring interactions \cite{thompson1982chaos}, bearing looseness \cite{muszynska1995chaotic} and the multi-degree-of-freedom impact oscillators \cite{yin2019complex},  may have abundant coexisting attractors when grazing occurs. Tiny differences in modelling will lead to different motion of the system { \cite{ing2010bifurcation,jeffrey2010catastrophic,qiu2019observer}.
For example, the motion of an impact oscillator experiences significant change due to a slight variation on its parameter when a grazing bifurcation is encountered \cite{ing2007experimental}. In \cite{nordmark1997universal}, Nordmark studied the characteristic scaling behaviour near grazing bifurcations, and used the self-similarity under scaling to derive a renormalised mapping. Nordmark  \cite{nordmark1991non} presented the grazing bifurcation of a hard impact oscillator, where the Poincar\'e map has a singular Jacobian, by using a first-order Taylor expansion. It has shown that the stability of the oscillator can be studied more precisely if its grazing events are computed more accurately. This paper will study a new method to improve the accuracy of calculating the grazing events by estimating the impacting moment of the system. Based on this accurate grazing trajectory, stability analysis of the system can be carried out.

In many applications \cite{stepan2006stability,beregi2019bifurcation,zhang2011dynamics,carvalho2018new,yan2017basins,qiu2019command,pyragas2019state} arise differential equations in which the derivative of the unknown functions at a certain time depends on the value of the function at previous time. These are so-called delay differential equations (DDEs). For example, Zhang \emph{et al}. \cite{zhang2011dynamics} studied a delayed pest control model which was a high-dimensional differential equation with impulsive effects at different fixed impulsive moments. In \cite{carvalho2018new}, Carvalho and Pinto used a mathematical model with delay to describe the dynamics of AIDS- related cancers with the treatment of HIV and chemotherapy. In \cite{yan2017basins}, Yan \emph{et al}. used the basin of a time-delayed system modelling cutting process to determine the unsafe cutting zone. The above studies are concerned with smooth DDEs. The analysis of nonsmooth DDEs is more challenging due to the lack of an accurate algorithm for computing the grazing events. Until now, there are very few systematic studies regarding to nonsmooth DDEs, which is the focus of this paper. The present work will study a new algorithm to determine the occurrence of grazing for improving computational accuracy and a new method for calculating Lyapunov exponents (LEs) along the trajectories of a nonsmooth DDE.

The LE of a trajectory is a quantity that characterises the rate of separation of infinitesimally near-by trajectories \cite{Mainzer12}. It determines a notion of sensitivity of this trajectory to perturbations in initial conditions. If the largest LE, which is referred to the maximal LE, is greater than zero,
any small perturbation of the initial condition will result in an exponential divergence of the resulting perturbed trajectory until the distance between the perturbed and unperturbed trajectories is no longer small. This sensitivity with respect to initial condition is one of the defining features of chaos. If the LE are identical for typical trajectories of an attractor in a dynamical system, one speaks of the LE for this attractor (or this dynamical system). The LE indicate predictability (or lack of it) for dynamical systems, such that they are considered as an important tool for studying the stability of dynamical systems. Therefore, the development of an efficient method for calculating the LEs of dynamical system is an active area of research, see e.g. \cite{parker2012practical,benettin1980lyapunov,wolf1985determining, dieci1997compuation,stefanski2000estimation, muller1995calculation,dellago1996lyapunov,jin2006method}. For finite-dimensional dynamical systems Benettin \emph{et al}. \cite{benettin1980lyapunov} introduced a systematic method for estimating the LEs of smooth dynamical systems. Wolf \emph{et al}. \cite{wolf1985determining} developed a method for extracting the largest LE  from an experimental time series. For nonsmooth systems, M{\"u}ller \cite{muller1995calculation} developed a model-based algorithm to calculate the LEs of nonlinear dynamical systems with discontinuities. They found that the required linearised equations must be supplemented by certain transition conditions when crossing the discontinuities. In \cite{dellago1996lyapunov}, Dellago \emph{et al}. generalised Benettin's classical algorithm and applied it to the case of dynamical systems where smooth streaming was interrupted by a differentiable map at discrete times. Lamba and Budd \cite{lamba1994scaling} have shown that the largest LE has a discontinuous jump at grazing bifurcations in Filippov systems and scales like $1/|\ln \epsilon|$, where $\epsilon$ is the bifurcation parameter. In contrast to ordinary differential equatrions (ODEs), DDEs are infinite dimensional systems. such that the computation of LEs for nonsmooth DDEs combines difficulties from discontinuities and high dimensionality. In principle, a DDE could be approximated by a high-dimensional ODE, which can be linearised along trajectories obtained by numerical integration \cite{farmer1982chaotic, chavez2020numerical}, such that the LEs can be constructed for the Poincar\'e map. Studies by Repin \cite{repin1965approximate} and Gy{\"o}ri and Turi \cite{gyori1991uniform} have shown that DDEs can be analysed using approximating high-dimensional ODEs. However, if the delay time is large, calculating the LEs of nonsmooth DDEs needs to store excessive history data points during delay period compared to smooth DDEs \cite{breda2006solution, breda2005pseudospectral, breda2012approximation}, e.g. the data at past encounters of the discontinuity. In this case the global convergence of the system cannot be guaranteed. Therefore, it may cause inaccuracy in calculating the eigenvalues of Jacobian matrix which is used for estimating the LEs of nonsmooth DDEs.

The contribution of the present work is the development of a novel method for precisely calculating the LEs of piecewise-smooth differential equations with a delayed argument, which can provide improved accuracy for stability analysis of periodic orbits. In detail, if an algorithm cannot estimate the point of discontinuity along trajectory with an accuracy of the same order as its integration method, especially in the grazing event, the expected discontinuous coefficients of the variational problem will have unexpectedly low accuracy leading to an accumulation of errors. Similar work was reported by M\"{u}ller \cite{muller1995calculation} who studied a method for constructing the map of systems with discontinuity, and combined it with the map obtained along the differentiable parts of the trajectories to generate a composition of Jacobian matrices for calculating LEs. However, M\"{u}ller's approach is difficult to implement for piecewise-smooth DDE due to its high dimension and complex dynamics, which could cause a high computational cost and an accumulation of computational errors at discontinuous moments. We address this issue in the present work, demonstrating our approach for the delayed piecewise-smooth oscillator. We construct a Poincar\'e map that consists of many local maps for each small time step, which are linearised for the LE computation. As the linearised Poincar\'e map requires accurate information about the time of crossing or grazing of a discontinuity (when impact occurs), we will introduce a grazing estimation algorithm to obtain an accurate Jacobian matrix for the oscillator. The novelty of our proposed method is that it can estimate the point of discontinuity locally along trajectories of piecewise-smooth DDEs, improving the accuracy of computations of the system trajectory and of the LEs. The proposed method can also be extended to other nonsmooth dynamical systems, such as the hard impact oscillator with a time-delayed controller or stick-slip vibrations with a delay term. To demonstrate the reliability of the method, we will carry out an error analysis for the nonzero eigenvalues of the Jacobian by adopting the spectral approximation methods introduced by Chatelin \cite{chatelin2011spectral} and Breda \emph{et al.} \cite{breda2006solution,breda2012approximation}. Our study indicates that the proposed method can reduce the error for the nonzero eigenvalues of the Jacobian by increasing the dimensions of the system of ODEs approximating the DDE slightly, which is generated by linearising the DDEs along trajectories obtained by numerical integration.

The rest of this paper is organised as follows. Section \ref{sec-approxddes} introduces the mathematical model of a periodically forced mechanical oscillator subjected to a one-sided soft impact. This is followed by some basic relevant definitions and preparations. Section \ref{sec-method} presents the method for constructing the Jacobian of Poincar\'e map of piecewise-smooth DDEs. However, such a construction is inaccurate due to the nonsmoothness of the considered system. Thus, Section \ref{modification-discontinuous} studies an estimation method for determining the points of discontinuity accurately. Here, two cases of grazing events are considered based on the geometry of the trajectory. Section \ref{convergence-analysis} uses linear operator theory to carry out an error analysis for the eigenvalues of the Jacobian, which can validate the reliability of our proposed method. In Section \ref{sec-computation}, the steps for computing LEs are detailed. Examples and several control scenarios of the oscillator are presented in Section \ref{sec-simulation} to demonstrate the accuracy of the method. Finally, some concluding remarks are drawn in Section \ref{sec-conclusion}.

\section{Mathematical model and relevant preparations}\label{sec-approxddes}

The impact oscillator shown in Fig.~\ref{fig-model} represents a mechanical system encountering intermittent so-called soft impacts, which will be studied in the present work. Soft impacts occur in mechanical systems when an object hits an obstacle of negligible mass but non-negligible stiffness. In Fig.~\ref{fig-model} the object is modelled by the block of mass $m$ and the obstacle is modelled by the spring with stiffness $k_2$ (a \emph{backlash} spring). The collision occurs when the distance $g$ between block and spring reaches $0$. Since at impact the spring is relaxed, the forces in the system depend continuously on $g$ (and, hence, on the position $y$ of the block), but the spring constants exerted by the backlash spring are discontinuous: $0$ for $g>0$, $k_2$ for $g=0$. Systems with soft impacts are common to a broad range of engineering applications, e.g. \cite{souza08,lazarek20,serdukova2020postgrazing,makarenkov12}, where the repeated collision of mechanical parts is unavoidable \cite{liu2017controlling}. The vibro-impact capsule system \cite{liu2013modelling,Chavez16} is a typical two-degrees-of-freedom dynamical system experiencing soft impacts and nonlinear friction. Any small variations in friction or system parameters (e.g. the stiffness of the \emph{backlash} spring) may lead to a qualitative change of the dynamics of the system \cite{liu2013friction,Liu17}. Thus, accurate prediction of its collision is crucial to fully understand the dynamics of the system, in particular, in the presence of time-delay effects \cite{Wojewoda08}.

The nondimensional equations of motion of the impact oscillator can
be written in a compact form as below \cite{ing2007experimental},
\begin{equation}\label{eq-imposc}
\begin{cases}
x'(\tau)=&v(\tau),\\
v'(\tau)=&a\omega^2\sin(\omega\tau)-2\zeta
v(\tau)-x(\tau)-\beta(x(\tau)-e)H(x(\tau)-e),
\end{cases}
\end{equation}
where $H(\cdot)$ stands for the Heaviside step function and $x'$,
$v'$ denote differentiation with respect to the nondimensional time
$\tau$.
The discontinuity boundary is fixed at $x=e$, with $e>0$ being the nondimensional gap to the rest point of the linear spring.
Eq.~\eqref{eq-imposc} was nondimensionalised from the representation in Fig.~\ref{fig-model} by introducing the following variables and parameters
\begin{equation*}\label{eq-nondim}
\setstretch{2.25}\begin{array}{r@{}lcr@{}lcr@{}lcr@{}l}
\omega_{n}=&\mbox{ }\sqrt{\dfrac{k_1}{m}}, & & \tau=&\mbox{ }\omega_{n}t, & & \omega=&\mbox{ }\dfrac{\Omega}{\omega_{n}}, & & \zeta=&\mbox{ }\dfrac{c}{2m\omega_{n}},\\
x=&\mbox{ }\dfrac{y}{y_{0}}, & & e=&\mbox{ }\dfrac{g}{y_{0}}, & &
a=&\mbox{ }\dfrac{A}{y_{0}}, & & \beta=&\mbox{
}\dfrac{k_{2}}{k_{1}},
\end{array}
\end{equation*}
where $y_{0}>0$ is an arbitrary reference distance, $\omega_{n}$ is
the natural angular frequency of the mass-spring system ($m$, $k_1$ in Fig.~\ref{fig-model}), $\omega$ is the ratio between forcing forcing frequency and natural frequency, $\beta$ is the stiffness ratio, $\zeta$ is the damping ratio, and $a$ is
the nondimensionalised forcing amplitude.

\begin{figure}[H]
\centering
\includegraphics[width=3in]{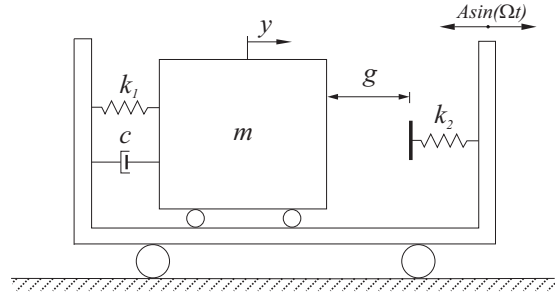}
\caption{Physical model of the soft impact oscillator \cite{liu2017controlling}.}\label{fig-model}
\end{figure}
In the present work, we will consider a control signal $u(\tau),\tau\ge0$, which will be superimposed on the system's external excitation as follows
\begin{equation}\label{eq-imposc-dde}
\begin{cases}
x'(\tau)=&v(\tau),\\
v'(\tau)=&\left(a\omega^2\sin (\omega\tau)+u(\tau)\right)-2\zeta
v(\tau)-x(\tau)-\beta(x(\tau)-e)H(x(\tau)-e),
\end{cases}
\end{equation}
where
\begin{equation}\label{eq-dde-u}
u(\tau)=k\left(v(\tau-\tau_{d})-v(\tau)\right),\mbox{ }\mbox{
}\mbox{ }\tau\geq0,
\end{equation}
defines the proportional feedback controller that feedbacks the difference between the current measurement of $v$ and a measurement of $v$ from some time $\tau_{d}$ ago \cite{pyragas1992continuous}.
In the expression above, $k\geq0$ represents the feedback gain of the controller
and $\tau_{d}>0$ stands for a predefined time delay. The control objective here is to avoid undesired chaotic responses and to suppress the multistability of the impact oscillator in the vicinity of the grazing events. We are interested in this type of time-delayed feedback, because it may result in a zero control signal if $\tau_d=2\pi/\omega$ and if \eqref{eq-dde-u} successfully stabilises a period-$1$ motion. This is the case even if we do not know the precise time profile of this period-$1$ motion, which is in contrast to standard linear feedback control $u(\tau)=k(v_\mathrm{ref}(\tau)-v(\tau)$. The asymptotically vanishing control signal is attractive in applications where energy consumption is a critical issue, e.g. \cite{Guo20}.

Eq.~(\ref{eq-imposc-dde}) can be rewritten in the form of a general piecewise continuous DDE with a periodic external excitation as
\begin{equation}\label{eq-hybsys-main}
\begin{cases}
\dot{y}(t)=f_{1}(y(t),y(t-\tau_{d}))+p(t), & ~\text{for}~ H(y(t),e)>0,\\
\dot{y}(t)=f_{2}(y(t),y(t-\tau_{d}))+p(t), & ~\text{for}~ H(y(t),e)<0,\\
y(t^{+})=y(t^{-}), & ~\text{for}~ H(y(t),e)=0,
\end{cases}
\end{equation}
where $f_{1,2}: \mathbb{R}^d\times  \mathbb{R}^d\rightarrow \mathbb{R}^d$, $H: \mathbb{R}^d\rightarrow  \mathbb{R}$ are
sufficiently smooth functions and $p: \mathbb{R}^{+}\to  \mathbb{R}^{d}$ is smooth and periodic with the period $T>0$. The delay $\tau_d$ is assumed to be positive but may be different from the period in general. In the present work, we only consider one single delay in the system for simplicity, and assume that for any $y, \bar{y}, y_{d}, \bar{y}_{d} \in \mathbb{R}^{d}$, $f_1$, $f_2$ and $H$ satisfy the Lipschitz condition
\begin{align*}
  |f_{1,2}(y,y_{d})-f_{1,2}(\bar{y},\bar{y}_{d}) |&\leq l_{1} |y-\bar{y} |+ l_{2}|y_{d}-\bar{y}_{d}  |\mbox{,}\\
  | H(y,e)-H(\bar{y},e)|&\leq l_{3}|y-\bar{y} |\mbox{,}
\end{align*}
where $l_{1},l_{2},l_3\ge 0$ and $|\cdot|$ is a norm on $\mathbb{R}^{d}$.
We assume that the initial condition is a suitable initial function on $[t_{0},t_{0}-\tau_{d}]$.
The general form \eqref{eq-hybsys-main} belongs to the class of \emph{hybrid dynamical systems} \cite{bernardo2008piecewise}, which consists of a flow (in our case only forward in time), combined with discrete events.

Take $N\in \mathbb{Z}^{+}$ sufficiently large, and define the discretisation grid points $\tau^{i}_{d}:=i\frac{\tau_{d}}{N}$, $i=0,\ldots,N$,  and $u_{i}(t):=y(t-\tau^{i}_{d})$ for all $t\ge 0$, $i=0,\ldots,N$. Eq.~(\ref{eq-hybsys-main}) can be approximated by a $d(N+1)$ dimensional piecewise-smooth discretised problem studied in \cite{repin1965approximate}, which will be presented in Section \ref{sec-method}.
This approximation method has also been studied by Krasovskii \cite{krasovskii1962analytic}, finding that the solution of the approximating system uniformly converges to the solution of the original DDEs when $N\to \infty$. By using the same approach, Gy\"ori and Turi \cite{gyori1991uniform} and Banks \cite{banks1979approximation} carried out convergence analyses for two DDEs. Breda \textit{et al.} \cite{breda2005pseudospectral} studied the characteristic roots of linear DDEs, and used a Runge-Kutta method to construct a high-dimensional approximating system. The nonzero eigenvalues of evolution operators were computed through a pseudospectral collection, which was used to analyse the asymptotic stability of DDEs. Since Eq.~(\ref{eq-hybsys-main}) is a piecewise-smooth DDE whose trajectories can encounter discontinuities, the methods used for smooth DDEs are not suitable, or, at least, converge with lower-than-expected order. Therefore, motivated by the periodic forcing of Eq.~(\ref{eq-hybsys-main}), our plan here is to derive a Poincar\'e map (also called stroboscopic map) for discretising the system and study linear stability of its orbits by considering the Jacobian matrix of the map in these orbits. After such a reduction to the Poincar\'e map, we will be able to define LEs for this time-discrete map.

For the piecewise DDE (\ref{eq-hybsys-main}), we consider a constant phase surface as the Poincar\'e section defined by
$
P^{T}_s:=\{ (y, t)\in  C([-\tau_d,0],\mathbb{R}^{d}) \times  \mathbb{R}^{+}|~ t=t_{0}+kT,~ k\in  \mathbb{Z}^{+}\}
$. For the corresponding Poincar\'e map
\begin{equation}\label{poin-map}
P:~P^{T}_s\to P^{T}_s
\end{equation}
the LEs can be defined as follows.

\begin{definition}\label{LEdis}\cite{parker2012practical}
For any initial condition $x_{0}\in P^T_s$, let $\{x_{m}\}^{\infty}_{m=0}$ be the corresponding orbit of the map $P$, and let $\lambda^{m}_{0 },\cdots,\lambda^{m}_{n}$ be the $n$ largest in modulus eigenvalues of $D P^{m}(x_{0})$, sorted such that $|\lambda^m_0|\geq\ldots\geq|\lambda^m_{n}|$. The Lyapunov exponents  of $x_{0}$ are
\begin{equation}\label{LEdis:formula}
\vartheta_{i}:=\lim_{m\to \infty}\ln |\lambda^{m}_{i}|^{\frac{1}{m}},i=1,\ldots,n
\end{equation}
whenever the limit exists for $x_0$ and for all $i\leq n$.
\end{definition}

The above definition is applicable to our map $P$ acting on the infinite dimensional space $P^T_s$, since $P$ is differentiable and its linearisation is bounded and has a spectrum consisting only of a sequence (finite or infinite) of eigenvalues of finite multiplicity converging to $0$ and zero. The expression in the limit \eqref{LEdis:formula} is not a practical recipe for computation since $\lambda^m_i$ may be very large or very small.

\section{Constructing the Jacobian matrix of the Poincar\'e map}\label{sec-method}
For the nonsmooth system with a delay $\tau_{d}$ smaller than its forcing period $T$, i.e. $0<\tau_{d}<T$, the period $T$ can be written as $T=n\tau_{d}+\Delta t$, for some $ n\in \mathbb{Z}^{+}$ and $\Delta t\in[0,\tau_{d})$. For any time interval $[t_{m},t_{m}+\tau_{d}]$, where $t_{m}=t_{1}+(m-1)T$, $t_{1}=t_{0}$ and $m\in \mathbb{Z}^{+}$, the solution of system (\ref{eq-hybsys-main}) can be approximated by $N$ steps of size $h=\frac{\tau_{d}}{N}$ by using numerical integration. The expression derived in this section initially ignore grazing of the discontinuity surface $\{H=0\}$. Section~\ref{modification-discontinuous} will explain how the expressions will be modified at the respective events. The modified Euler integration formula \cite{stoer2013introduction} gives for a single step of size $h=\tau_d/N$
\begin{align}\label{numinteg}
u_{0}(t_{m}+h)=&\,u_{0}(t_{m})+\tfrac{h}{2}\big[f_{j}(u_{0}(t_{m}),u_{0}(t_{m}-hN)) \nonumber \\
&+f_{j}(u_{0}(t_{m}+h),u_{0}(t_{m}-h(N-1)))\big]
+\tfrac{h}{2}\big[p(t_{m})+p(t_{m}+h)\big],
\end{align}
(here written only for the first step at $t_m$) where
\begin{equation*}
\begin{cases}
j=1, & ~\text{if}~ H(u_{0}(t_{m}),e)>0,\\
j=2, & ~\text{if}~ H(u_{0}(t_{m}),e))<0,\\
u_{0}(t^{+}_{m})=u_{0}(t^{-}_{m}), & ~\text{if}~ H(u_{0}(t_{m}),e)=0.
\end{cases}
\end{equation*}

Iterating this map $N+1$ times gives a discretised map for the delay-time interval $[t_{m},t_{m}+\tau_{d}]$, which we call $P_{d}:\mathbb{R}^{d(N+1)}\rightarrow \mathbb{R}^{d(N+1)}$. It satisfies
\begin{equation}\label{first-local-map}
U_{m,1}=P_{d}(U_{m,0}),
\end{equation}
where $U_{m,0}: =(u_{N}^{T}(t_{m}),\cdots,u_{1}^{T}(t_{m}),u_{0}^{T}(t_{m}))^{T}\in \mathbb{R}^{d(N+1)}$ and $U_{m,1}: =(u_{N}^{T}(t_{m}+\tau_{d}),\cdots,u_{1}^{T}(t_{m}+\tau_{d}),u_{0}^{T}(t_{m}+\tau_{d}))^{T} \in \mathbb{R}^{d(N+1)}$, and we use the general convention that $u_i(t)=u_0(t-(i/N)\tau_d)$ for arbitrary $i\in\{0,\ldots,N\}$ and $t$}.
Iterating the map $P_d$ $n$ times, we can obtain a map $P_d^n$ from $U$ at time $t_m$ to $U$ at time $t_m+n\tau_d$,
\begin{equation}\label{combmap}
U_{m,n}=P_{d}\circ\cdots\circ P_{d}(U_{m,0})=P^{n}_{d}(U_{m,0}),
\end{equation}
where $U_{m,i}:=(u_{N}^{T}(t_{m}+ihN),\cdots,u_{0}^{T}(t_{m}+ihN))^{T} \in \mathbb{R}^{d(N+1)}$.
Finally the discretised map for the time $\Delta t$ is defined as $P_{\Delta t}:\mathbb{R}^{d(N+1)}\rightarrow \mathbb{R}^{d(N+1)}$, which can be represented as
\begin{equation}\label{smallmap}
U_{m,n+\Delta N}=P_{\Delta t}(U_{m,n}),
\end{equation}
where $U_{m,n+\Delta N}:=(u_{N}^{T}(t_{m}+h(nN+\Delta N))^{T},\cdots,u_{0}^{T}(t_{m}+h(nN+\Delta N))^{T} \in \mathbb{R}^{d(N+1)}$ and $\Delta N:=\frac{\Delta t}{h}$. 
Thus combining Eqs.~(\ref{combmap}) and (\ref{smallmap}) we can construct map $P_\mathrm{disc}$ as the discretised Poincar\'e map $P$ advancing by time $T$
\begin{equation}\label{mainmap}
U_{m, n+\Delta N}=P_\mathrm{disc}(U_{m+1,0})=P_{\Delta t} \circ P^{n}_{d} (U_{m,0}),
\end{equation}
which can then iterate further by setting $U_{m+1,0}=U_{m,n+\Delta N}$. For an arbitrary perturbation $\delta U$ is applied, the variational equation for $P_\mathrm{disc}$ can be written as
\begin{equation}\label{syspds}
\delta U_{m+1,0}=\sum^{N+1}_{i=1}\frac{\partial P_\mathrm{disc}(U_{m,0})}{\partial u_{i-1}(t_{m})}\delta u_{i-1}(t_{m}),
\end{equation}
where $\delta U_{m,0}:=(\delta u_{N}^{T}(t_{m}),\cdots,\delta u_{1}^{T}(t_{m}),\delta u_{0}^{T}(t_{m}))^{T} \in \mathbb{R}^{d(N+1)}$, and we use again the convention that $\delta u_{i}(t):=\delta u(t-\tau_d^i), i=0,\cdots,N$. In fact, Eq.~(\ref{syspds}) can be obtained from discretising the continuous variational equation of system (\ref{eq-hybsys-main}), and its form can be obtained as
\begin{equation}\label{varia-o}
\frac{\mathrm{d} }{\mathrm{d}t}\delta u_{0}(t)=\frac{\partial f_{j}(t,u_{0}(t),u_{N}(t))}{\partial u_{0}}\delta u_{0}(t)
+\frac{\partial f_{j}(t,u_{0}(t),u_{N}(t))}{\partial u_{N}}\delta u_{N}(t),
\end{equation}
where
\begin{equation*}
\begin{cases}
j=1, & ~\text{if}~ H(u_{0}(t),e)>0,\\
j=2, & ~\text{if}~ H(u_{0}(t),e)<0,\\
u_{0}(t^{+})=u_{0}(t^{-}), & ~\text{if}~ H(u_{0}(t),e)=0.
\end{cases}
\end{equation*}
An example initial function $\phi_{\delta }$ for \eqref{varia-o} is of the form $\phi_{\delta }(t_{1})=(\epsilon,0,\cdots,0)^{T}\in  \mathbb{R}^{d}$ and $\phi_{\delta }(t)=(0,\cdots,0)^{T}\in  \mathbb{R}^{d}$ for $t\in [t_{1}-\tau_{d},t_{1})$, and sufficiently small $\epsilon$. 
Discretising Eq.~(\ref{varia-o}) in the interval $[t_{m},t_{m}+n\tau_{d}]$ by using the modified Euler integration gives
\begin{align}\label{disvs}
\delta u_{0}(t_{m}+lh)
=&\,\delta u_{0}(t_{m}+(l-1)h) \nonumber \\
&+\tfrac{h}{2}\big[A_{m,l}\delta u_{0}(t_{m}+(l-1)h+B_{m,l}\delta u_{0}(t_{m}-(N-l+1)h)]  \\
&+\tfrac{h}{2}[A_{m,l+1}\delta u_{0}(t_{m}+lh)+B_{m,l+1}\delta u_{0}(t_{m}-(N-l)h)\big], \nonumber
\end{align}
where $l=1,\cdots,N,\cdots,nN+\Delta N$, $A_{m,l}=\frac{\partial f_{j}(u_{0}(t),u_{N}(t))}{\partial u_{0}}|_{t=t_{m}+h(l-1)}$, $B_{m,l}=\frac{\partial f_{j}(u_{0}(t),u_{N}(t))}{\partial u_{N}}|_{t=t_{m}+h(l-1)}$ and $m\in \mathbb{Z}^{+}$.
Rewriting Eq.~(\ref{disvs}) in a matrix form gives
\begin{align}\label{matrixfuncs}
\begin{bmatrix}
  \delta u_{N}({t_{m}+lh})  \\
  \vdots \\
  \delta u_{1}({t_{m}}+lh)\\
  \delta u_{0}({t_{m}}+lh)
\end{bmatrix}
&=M_{m,l}\begin{bmatrix}
  \delta u_{N}({t_{m}+(l-1)h})  \\
    \vdots \\
    \delta u_{1}({t_{m}+(l-1)h})  \\
  \delta u_{0}({t_{m}}+(l-1)h)
  \end{bmatrix} ,
\end{align}
where
\begin{align*}
M_{m,l}=\hat{M}_{m,l} \tilde{M}_{m,l},
\end{align*}
\begin{align*}
\hat{M}_{m,l}=\left[\begin{array}{cccc}
  I & \cdots & 0 & 0 \\
  \vdots & \ddots & \vdots &  \vdots \\
  0&  \cdots& I & 0 \\
  -\tfrac{h}{2}B_{m,l+1}&\cdots& 0  & I-\tfrac{h}{2}A_{m,l+1}\\
\end{array}\right]^{-1},
\end{align*}
and
\begin{align*}
\tilde{M}_{m,l}=\left[\begin{array}{cccc}
  0 & I &\cdots & 0 \\
  \vdots & \vdots& \ddots & \vdots \\
  0&0&  \cdots&I \\
 \tfrac{h}{2} B_{m,l}& 0 &\cdots & I+\tfrac{h}{2}A_{m,l}\\
\end{array}\right].
\end{align*}

By using the map (\ref{first-local-map}), the matrix form of the variational equation (\ref{matrixfuncs}) can be rewritten as
\begin{equation*}
\delta U_{m,N}=M_{m,N}\circ\cdots\circ M_{m,1}\delta U_{m,0}.
\end{equation*}
Since we have $n$ maps, combining all the maps for the interval $ [t_{m},t_{m}+T] $ gives
\begin{equation*}
\delta U_{m,n}=M_{m,nN}\circ\cdots\circ M_{m,2}\circ M_{m,1}\delta U_{m,0}.
\end{equation*}
In addition, the map $P_{\triangle t}$ for the interval $[t_{m}+n\tau_{d},t_{m}+T]$ can be written as
\begin{align}\label{m-mainvar}
\delta U_{m+1,0}=M_{m,nN+\Delta N}\circ\cdots\circ M_{m,nN}\delta U_{m,n}.
\end{align}
Finally, the overall variational equation can be obtained as
\begin{equation} \label{mainvar}
\delta U_{m+1,0}=M_{m}\delta U_{m,0},
\end{equation}
where $M_{m}=M_{m,nN+\Delta N}\circ \cdots\circ M_{m,nN}\circ\cdots\circ M_{m,1}$ is the approximation of Jacobian matrix of the Poincar\'e map $P$.

Similarly, for the system with a large delay time, e.g. $\tau_{d} \ge T$, the solution of system (\ref{eq-hybsys-main}) can be approximated by $N$ steps f size $h=\frac{\tau_{d}}{N}$ by using numerical integration, which can be considered as a special case of the nonsmooth system with a small delay time ($0<\tau_{d}<T$) when $n=0$.
Let $N_{T}=\frac{T}{h}$ be the sample number for one period $T$, construct the map $P_{d}$, and combine all the linearised maps at the interval $[t_{m},t_{m}+T]$.
Finally, we can obtain the same variational equation as Eq.~(\ref{mainvar}) and the Jaocbian matrix of the Poincar\'e map $P$.

\section{Modifying the algorithm at the discontinuity}\label{modification-discontinuous}
\begin{figure}[h!]
\centering
\includegraphics[width=0.55\textwidth]{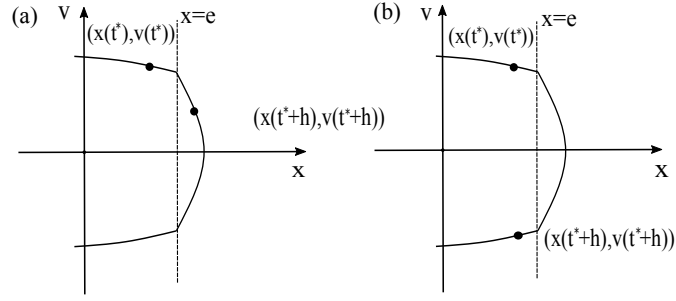}
\caption{(a) Case 1: for $t=t^{*}>0$, such that $H_{1}:=H(u_{0}(t^{*}),e)<0$ and $H_{2}:=H(u_{0}(t^{*}+h),e)>0$ (or $H_{1}>0$ and $H_{2}<0$).
(b) Case 2: for $t=t^{*}>0$, and there exists $\delta t\in (0,h)$, such that $H_{1}:=H(u_{0}(t^{*}),e)< 0$, $H_{2}:=H(u_{0}(t^{*}+h))<0$ and $H_{cr,1}:=H(u_{0}(t^{*}+\delta t),e)=0$ (or $H_{1}>0$, $H_{2}>0$ and $H_{cr,1}=0$).}\label{fig-grazing}
\end{figure}

In this section, we will discuss a special phenomenon of the impact oscillator, the so-called crossing and grazing events. Since the system has rich complex dynamics when it experiences grazing \cite{ing2010bifurcation,jiang2016geometrical}, a careful consideration in calculating this discontinuous moment is required. In addition, the global error of our proposed algorithm will depend on how accurately we capture the effect of switching, as the error made at the switching boundary could accumulate, leading to unexpected large global error. Therefore, during the grazing event, we need to modify our proposed algorithm from Section~\ref{sec-method} by considering the two grazing cases illustrated in Fig.~\ref{fig-grazing}.

\subsection{Case 1}
For Case 1, we assume that for time step $l^{*}\in \mathbb{Z}^{+}$ at time $t^{*}:=t_{m}+(l^{*}-1)h$ the switching function $H$ changes sign:$H_{1}:=H(u_{0}(t^{*}),e)<0$
and $H_{2}:=H(u_{0}(t^{*}+h),e)>0$, or $H_{1}>0$, $H_{2}<0$. Thus, we expect that for some time $\delta t\in(0,h)$, the switching fucntion is zero: $H_{cr,1}:=H(u(t^{*}+\delta t),e)=0$. In order to guarantee the order of convergence of our proposed algorithm to $O(h^{2})$, the crossing time $\delta t$ needs to be estimated first. Since $\delta t<h$, the condition $H_{cr,1}=0$ can be linearised as
\begin{equation*}
H_{cr,1}\approx H(u(t^{*})+\dot u(t^{*})\delta t,e)\approx H_{1}+\tfrac{d}{du}H_{1}[\dot u(t^{*})\delta t]=0,
\end{equation*}
such that
\begin{equation}\label{estimation-time}
\delta t=\frac{-H_{1}}{\tfrac{d}{du}H_{1}  [\dot u(t^{*})]}.
\end{equation}

Once $\delta t$ is calculated, the switching time $t^{*}+\delta t$ can be obtained, and the variational equation at the step crossing the switching can be written as
\begin{align}\label{disvs-impact}
\delta u_{0}(t^{*}+\delta t)=&\,\delta u_{0}(t^{*})+\tfrac{\delta t}{2}[A_{m,l^{*}}\delta u_{0}(t^{*})+B_{m,l^{*}}\delta u_{N}(t^{*})] \nonumber \\
&+\tfrac{\delta t}{2}[A^{\delta t}_{m, l^{*}}\delta u_{0}(t^{*}+\delta t)+B^{\delta t}_{m,l^{*}}\delta u_{N}(t^{*}+\delta t)],
\end{align}
where
$A^{\delta t}_{m,l^{*}}=\frac{\partial f_{j}(u_{0}(t),u_{N}(t))}{\partial u_{0}}|_{t=t^{-}_{m}+h(l^{*}-1)+\delta t}$, $B^{\delta t}_{m,l^{*}}=\frac{\partial f_{j}(u_{0}(t),u_{N}(t))}{\partial u_{N}}|_{t=t^{-}_{m}+h(l^{*}-1)+\delta t}$ and $l^{*}=1,\cdots,N,$ $\cdots,nN+\Delta N$.
Thus the discretied map from $t^{*}$ to $t^{*}+\delta t$ can be written as
\begin{align}\label{matrixfuncs-ca1-1}
\begin{bmatrix}
  \delta u_{N}(t^{*}+\delta t)  \\
  \vdots \\
  \delta u_{1}(t^{*}+\delta t)\\
  \delta u_{0}(t^{*}+\delta t)
\end{bmatrix}
&=M^{\delta t}_{m,l^{*}}\begin{bmatrix}
  \delta u_{N}(t^{*})  \\
    \vdots \\
    \delta u_{1}(t^{*})  \\
  \delta u_{0}(t^{*})
  \end{bmatrix} ,
\end{align}
where
\begin{align*}
M^{\delta t}_{m,l^{*}}=\hat{M}^{\delta t}_{m,l^{*}} \tilde{M}^{\delta t}_{m,l^{*}}  ,
\end{align*}
\begin{align*}
\hat{M}^{\delta t}_{m,l^{*}}:=\left[\begin{array}{cccc}
  I & \cdots & 0 & 0 \\
  \vdots & \ddots & \vdots &  \vdots \\
  0&  \cdots& I & 0 \\
  -\frac{\delta t}{2}B^{\delta t}_{m,l^{*}}&\cdots& 0  & I-\frac{\delta t}{2}A^{\delta t}_{m,l^{*}}\\
\end{array}\right]^{-1},
\end{align*}
\begin{align*}
\tilde{M}^{\delta t}_{m,l^{*}} :=\left[\begin{array}{cccc}
  0 & I &\cdots & 0 \\
  \vdots & \vdots& \ddots & \vdots \\
  0&0&  \cdots&I \\
 \frac{\delta t}{2}B_{m,l^{*}}& 0 &\cdots & I+\frac{\delta t}{2}A_{m,l^{*}}\\
\end{array}\right],
\end{align*}
$A_{m,l^{*}}=\frac{\partial f_{j}(u_{0}(t),u_{N}(t))}{\partial u_{0}}|_{t=t_{m}+h(l^{*}-1)}$ and $B_{m,l^{*}}=\frac{\partial f_{j}(u_{0}(t),u_{N}(t))}{\partial u_{N}}|_{t=t_{m}+h(l^{*}-1)}$.
It is worth noting that $\delta u_{i}(t^{*}+\delta t)$ can be approximated through linear interpolation based on the historical data obtained from the delayed time interval which also includes the grazing data.

Similarly, for the time interval $[t^{*}+\delta t, t^{*}+h]$, we can obtain
\begin{align}\label{matrixfuncs-ca1-2}
\begin{bmatrix}
    \delta u_{N}(t^{*}+h)  \\
    \vdots \\
    \delta u_{1}(t^{*}+h)  \\
  \delta u_{0}(t^{*}+h)
  \end{bmatrix}
&=\bar{M}^{h}_{m,l^{*}}\begin{bmatrix}
  \delta u_{N}(t^{*}+\delta t)  \\
  \vdots \\
  \delta u_{1}(t^{*}+\delta t)\\
  \delta u_{0}(t^{*}+\delta t)
    \end{bmatrix} ,
\end{align}
where $\bar{M}^{h}_{m,l^{*}}:=\hat{M}^{h}_{m,l^{*}} \tilde{M}^{h}_{m,l^{*}} $,
\begin{align*}
\tilde{M}^{h}_{m,l^{*}} :=\left[\begin{array}{cccc}
  0 & I &\cdots & 0 \\
  \vdots & \vdots& \ddots & \vdots \\
  0&0&  \cdots&I \\
 \frac{h-\delta t}{2}B^{\delta t}_{m,l^{*}}& 0 &\cdots & I+\frac{h-\delta t}{2}A^{\delta t}_{m,l^{*}}\\
\end{array}\right]
\end{align*}
and
\begin{align*}
\hat{M}^{h}_{m,l^{*}}:=\left[\begin{array}{cccc}
  I & \cdots & 0 & 0 \\
  \vdots & \ddots & \vdots &  \vdots \\
  0&  \cdots& I & 0 \\
  -\frac{h-\delta t}{2}B_{m,l^{*}+1}&\cdots& 0  & I-\frac{h-\delta t}{2}A_{m,l^{*}+1}\\
\end{array}\right]^{-1}.
\end{align*}

Finally, we have
 \begin{align}\label{matrixfuncs-ca1}
\begin{bmatrix}
    \delta u_{N}(t^{*}+h)  \\
    \vdots \\
    \delta u_{1}(t^{*}+h)  \\
  \delta u_{0}(t^{*}+h)
  \end{bmatrix}
&=\bar{M}^{h}_{m,l^{*}} M^{\delta t}_{m,l^{*}}\begin{bmatrix}
  \delta u_{N}(t^{*})  \\
  \vdots \\
  \delta u_{1}(t^{*})\\
  \delta u_{0}(t^{*})
    \end{bmatrix}.
\end{align}
Therefore, when Case 1 occurs,
$\bar{M}^{h}_{m,l^{*}} M^{\delta t}_{m,l^{*}}$ should be inserted between $M_{m,l^{*}+1}$ and $M_{m,l^{*}}$ for the time interval $[t^{*},t^{*}+h]$ in Eq.~(\ref{m-mainvar}). The expressions for crossing events from $H>0$ to $H<0$ look identical, except that the subscripts $1$ and $2$ for $f$ are reversed.

\subsection{Case 2}
Let $\delta t$ be the first crossing time for Case 2, which can be calculated based on Eq.~(\ref{estimation-time}). We define $\delta t^{*}$ as the time where $H$ is maximal, such that $H_{\max}:=H(u(t^{*}+\delta t+\delta t^{*}))=\max_{t\in [t^{*},t^{*}+h]}H(u(t),e)$, and $\delta\bar{t}$ as the time where $H$ changes sign back, such that $H_{cr,2}:=H(u(t^{*}+\delta t_{g}))=0$, where $\delta t_{g}:=\delta t+\delta t^{*}+\delta\bar{t}$.
The estimate of $\delta t$ follows Eq.~(\ref{estimation-time}).
From a computational point of view, Case 2 can be triggered either by
(i) $H_{1}<0$, $H_{2}>0$, $\tfrac{d}{dt}H_{1}>0$, $\tfrac{d}{dt}H_{2}>0$ and $0<\delta t_{g}<h$,
or (ii) $H_{1}>0$, $H_{2}<0$, $\tfrac{d}{dt}H_{1}<0$, $\tfrac{d}{dt}H_{2}<0$ and $0<\delta t_{g}<h$.

Since 
\begin{align*}
\tfrac{d}{dt}H(t^*+\delta t+t)\vert_{t=\delta t^*}
\approx& \tfrac{d}{du}H_{cr,1} [\dot u(t^{*}+\delta t)+\ddot u(t^{*}+\delta t)\delta t^{*}]+\tfrac{d^{2}}{du^{2}}H_{cr,1}[\dot u^{2}(t^{*}+\delta t)\delta t^{*}]=0,
\end{align*}
we have
\begin{align}\label{case-2-middle}
\delta t^{*}
=\frac{-\tfrac{d}{du}H_{cr,1}[\dot u(t^{*}+\delta t)]}{
\tfrac{d}{du}H_{cr,1}[\ddot u(t^{*}+\delta t)]+\tfrac{d^{2}}{du^{2}}H_{cr,1} [\dot u^{2}(t^{*}+\delta t)]}.
\end{align}
For $\delta \bar t$ we have
\begin{align*}
H_{cr,2}\approx&\,H_{\max}+\tfrac{d}{du} H_{\max} [ \dot u(t^{*}+\delta t+\delta t^{*})] \delta \bar{t} \\
\approx&\,H_{\mathrm{cros},1}+\tfrac{d}{du} H_{\mathrm{cros},1} [\dot u(t^{*}+\delta t)]\delta t^{*}\\
&+\big[ \tfrac{d}{du}H_{\mathrm{cros},1}+ \tfrac{d^{2}}{du^{2}}H_{\mathrm{cros},1} [\dot u(t^{*}+\delta t)\delta t^{*}]\big] \big[\dot u(t^{*}+\delta t)+\ddot u(t^{*}+\delta t) \delta t^{*} \big]\delta \bar{t}=\,0,
\end{align*}
which gives
\begin{align}\label{case-2-end}
\delta\bar{ t}=&-\big[H_{cr,1}+\tfrac{d}{du} H_{cr,1} [\dot u(t^{*}+\delta t)]\delta t^{*}\big]  \big[ \tfrac{d^{2}}{du^{2}}H_{cr,1}\nonumber\\
&+\tfrac{d^{2}}{du^{2}}H_{cr,1} [\dot u(t^{*}+\delta t)\delta t^{*}]\big] \big[\dot u(t^{*}+\delta t)+\ddot u(t^{*}+\delta t) \delta t^{*} \big]^{-1}.
\end{align}
Therefore, for the step from $t^{*}$ to $t^{*}+\delta t$, the variational equation can be written as
\begin{align}\label{matrixfuncs-ca2-1}
\begin{bmatrix}
  \delta u_{N}(t^{*}+\delta t)  \\
  \vdots \\
  \delta u_{1}(t^{*}+\delta t)\\
  \delta u_{0}(t^{*}+\delta t)
\end{bmatrix}
&=M^{\delta t}_{m,l^{*}}\begin{bmatrix}
  \delta u_{N}(t^{*})  \\
    \vdots \\
    \delta u_{1}(t^{*})  \\
  \delta u_{0}(t^{*})
  \end{bmatrix}.
\end{align}
For the step from $t^{*}+\delta t$ to $t^{*}+\delta t_{\mathrm{graz}}]$ we have
\begin{align}\label{matrixfuncs-ca2-end}
\begin{bmatrix}
    \delta u_{N}(t^{*}+\delta t_{\mathrm{graz}})  \\
    \vdots \\
    \delta u_{1}(t^{*}+\delta t_{\mathrm{graz}})  \\
  \delta u_{0}(t^{*}+\delta t_{\mathrm{graz}})
  \end{bmatrix}
&=M^{\delta t_{\mathrm{graz}}}_{m,l^{*}}\begin{bmatrix}
  \delta u_{N}(t^{*}+\delta t)  \\
  \vdots \\
  \delta u_{1}(t^{*}+\delta t)\\
  \delta u_{0}(t^{*}+\delta t)
    \end{bmatrix} ,
\end{align}
where
\begin{align*}
M^{\delta t_{\mathrm{graz}}}_{m,l^{*}}:=\hat{M}^{\delta t_{\mathrm{graz}}}_{m,l^{*}} \tilde{M}^{\delta t_{\mathrm{graz}}}_{m,l^{*}}  ,
\end{align*}

\begin{align*}
\hat{M}^{\delta t_{\mathrm{graz}}}_{m,l^{*}}:=&
\left[\begin{array}{cccc}
  I & \cdots & 0 & 0 \\
  \vdots & \ddots & \vdots &  \vdots \\
  0&  \cdots& I & 0 \\
  -\frac{\delta t^{*}+\bar{\delta t}}{2}B^{\delta t_{\mathrm{graz}}}_{m,l^{*}}&\cdots& 0  & I-\frac{\delta t^{*}+\bar{\delta t}}{2}A^{\delta t_{\mathrm{graz}}}_{m,l^{*}}\\
\end{array}\right]^{-1},
\end{align*}

\begin{align*}
 \tilde{M}^{\delta t_{\mathrm{graz}}}_{m,l^{*}} :=\left[\begin{array}{cccc}
  0 & I &\cdots & 0 \\
  \vdots & \vdots& \ddots & \vdots \\
  0&0&  \cdots&I \\
 \frac{\delta t^{*}+\bar{\delta t}}{2}B^{\delta t}_{m,l^{*}}& 0 &\cdots & I+\frac{\delta t^{*}+\bar{\delta t}}{2}A^{\delta t}_{m,l^{*}}\\
\end{array}\right],
\end{align*}
$A^{\delta t_{\mathrm{graz}}}_{m,l^{*}}=\frac{\partial f_{j}(u_{0}(t),u_{N}(t))}{\partial u_{0}}|_{t=t^{-}_{m}+h(l^{*}-1)+\delta t_{\mathrm{graz}}}$ and $B^{\delta t_{\mathrm{graz}}}_{m,l^{*}}=\frac{\partial f_{j}(u_{0}(t),u_{N}(t))}{\partial u_{N}}|_{t=t^{-}_{m}+h(l^{*}-1)+\delta t_{\mathrm{graz}}}$.
For the period $ [t^{*}+\delta t_{\mathrm{graz}},t^{*}+h]$,
\begin{align}\label{matrixfuncs-ca2-3}
\begin{bmatrix}
    \delta u_{N}(t^{*}+h)  \\
    \vdots \\
    \delta u_{1}(t^{*}+h)  \\
  \delta u_{0}(t^{*}+h)
  \end{bmatrix}
&=\bar{M}^{h}_{m,l^{*}}\begin{bmatrix}
    \delta u_{N}(t^{*}+\delta t_{\mathrm{graz}})  \\
    \vdots \\
    \delta u_{1}(t^{*}+\delta t_{\mathrm{graz}})  \\
  \delta u_{0}(t^{*}+\delta t_{\mathrm{graz}})
    \end{bmatrix} ,
\end{align}
where
\begin{align*}
\bar{M}^{h}_{m,l^{*}}=\hat{M}^{h}_{m,l^{*}} \tilde{M}^{h}_{m,l^{*}}  ,
\end{align*}
\begin{align*}
\hat{M}^{h}_{m,l^{*}}:=&\left[\begin{array}{cccc}
  I & \cdots & 0 & 0 \\
  \vdots & \ddots & \vdots &  \vdots \\
  0&  \cdots& I & 0 \\
  -\frac{h-\delta t_{\mathrm{graz}}}{2}B_{m,l^{*}+1}&\cdots& 0  & I-\frac{h-\delta t_{\mathrm{graz}}}{2}A_{m,l^{*}+1}\\
\end{array}\right]^{-1}
\end{align*}
and
\begin{align*}
 \tilde{M}^{h}_{m,l^{*}}:=&\left[\begin{array}{cccc}
  0 & I &\cdots & 0 \\
  \vdots & \vdots& \ddots & \vdots \\
  0&0&  \cdots&I \\
 \frac{h-\delta t_{\mathrm{graz}}}{2}B^{\delta t_{\mathrm{graz}}}_{m,l^{*}}& 0 &\cdots & I+\frac{h-\delta t_{\mathrm{graz}}}{2}A^{\delta t_{\mathrm{graz}}}_{m,l^{*}}\\
\end{array}\right].
\end{align*}
Finally, we have
 \begin{align}\label{matrixfuncs-ca2}
\begin{bmatrix}
    \delta u_{N}(t^{*}+h)  \\
    \vdots \\
    \delta u_{1}(t^{*}+h)  \\
  \delta u_{0}(t^{*}+h)
  \end{bmatrix}
&=\bar{M}^{h}_{m,l^{*}}M^{\delta t_{\mathrm{graz}}}_{m,l^{*}}M^{\delta t}_{m,l^{*}}\begin{bmatrix}
  \delta u_{N}(t^{*})  \\
  \vdots \\
  \delta u_{1}(t^{*})\\
  \delta u_{0}(t^{*})
    \end{bmatrix} ,
\end{align}
Thus, once Case 2 is encountered, $\bar{M}^{h}_{m,l^{*}}M^{\delta t_{\mathrm{graz}}}_{m,l^{*}}M^{\delta t}_{m,l^{*}}$ should be inserted between $M_{m,l^{*}+1}$ and $M_{m,l^{*}}$ in Eq.~(\ref{m-mainvar}) for the step from $t^{*}$ to $t^{*}+h$.

From the discussion above, we can obtain an accurate Jacobian matrix for the Poincar\'e map  (\ref{poin-map}). In the next section, we will discuss the convergence of eigenvalues of the Jacobian matrix when a perturbation is introduced in order to ensure the accuracy of our proposed method.

\section{Convergence analysis}\label{convergence-analysis}

\subsection{Properties of the evaluation operator}
According to \cite{chatelin1973convergence,chatelin2011spectral}, the spectrum of the Jacobian for the Poincar\'e map consists of eigenvalues and $0$.
So we will study the Poincar\'e map of Eq.~(\ref{varia-o}) and its relevant Jacobian.

For the space $\mathbb{C}^{d}$, assume $\mathbb{P}:=[t_{1},t_{1}+\Delta T]$, which is an bounded interval of $\mathbb{R}$ and $\Delta T<+\infty$. $C(\mathbb{P}, \mathbb{C}^{d})$ denotes the Banach space with all bounded continuous functions from $\mathbb{P}$ to $\mathbb{C}^{d}$ with the norm $||u||_{C}=\max_{t\in \mathbb{P}}|u(t)|$, where $u\in C(\mathbb{P}, \mathbb{C}^{d})$ and $|\cdot|$ is a given norm on $\mathbb{C}^{d}$.

Now, we rewrite Eq.~(\ref{varia-o}) as 
\begin{equation}\label{abstact-vari}
\begin{cases}
\frac{\mathrm{d}}{\mathrm{d}t}\delta u_{0}(t)=F(t,\delta u_{0}(t),\delta u_{N}(t)),\quad\text{where}\ t\in\mathbb{P}\ \mathrm{and}\ F:\mathbb{P}\times \mathbb{C}^{d}\times\mathbb{C}^{d}\to\mathbb{C}^{d},\\
\delta u_{0}(t)=\phi_{\delta}\left(t\right),\quad\text{where}\ t\in[t_{1}-\tau_{d},t_{1}]\ \mathrm{and}\ \phi_{\delta }\in C([t_{1}-\tau_{d},t_{1}],\mathbb{C}^d),
\end{cases}
\end{equation}
where $\phi_{\delta }$ is defined in Eq.~(\ref{varia-o}).
Here, we assume $\delta u_{d}(t)=\delta u_{N}(t)$, and $F$ can be written as
\begin{align}\label{linear-property}
F(t,\delta u_{0}(t),\delta u_{d}(t))=
F_{j,1}(t)\delta u_{0}(t)+F_{j,2}(t)\delta u_{d}(t),
\end{align}
where
\begin{equation*}
\begin{cases}
j=1, ~\text{if}~ H(u_{0}(t),e)>0,\\
j=2, ~\text{if}~ H(u_{0}(t),e)<0,\\
F(t^{-},\delta u_{0}(t^{-}),\delta u_{d}(t^{-}))=F(t^{+},\delta u_{0}(t^{+}),\delta u_{d}(t^{+})), ~\text{if}~ H(u_{0}(t),e)=0,
\end{cases}
\end{equation*}
$F_{j,1}(t):=\frac{\partial f_{j}(t,u_{0}(t),u_{d}(t))}{\partial u_{0}}$, and $F_{j,2}(t):=\frac{\partial f_{j}(t,u_{0}(t),u_{d}(t))}{\partial u_{d}}$.

According to \cite{breda2012approximation}, nonautonomous delayed dynamical system can be represented as an evolution operator.
So, for any $t_{1}\in \mathbb{P}$ and sufficiently small $h>0$, we have
\begin{equation}\label{evolution-opera}
U(t_{1}+h,t_{1})\phi_{\delta }=\delta u_{0}(t_{1}+h),
\end{equation}
where $\delta u_{0}(t_{1}+h)$ is the solution of Eq.~(\ref{abstact-vari}) at $t=t_{1}+h$.
For any time $t=t_{1}+N_{t}h, ~\forall N_{t}\in \mathbb{Z}^{+}$, $\delta u_{0}(t)$ can be written as
\begin{align*}
\delta u_{0}(t)=&\, U(t_{1}+hN_{t},t_{1}+h(N_{t}-1))\cdots U(t_{1}+2h,t_{1}+h) U(t_{1}+h,t_{1})\phi_{\delta }.
\end{align*}

Next, we will construct the approximation operator with finite dimension for the evolution operator $U(t_{1}+h,t_{1})$.
In order to simplify our discussion, we define the following spaces
\begin{equation*}
\mathcal{P}:=C([t_{1}-\tau_{d},t_{1}],\mathbb{C}^{d}),
\end{equation*}
and
\begin{equation*}
\mathcal{P}^{+}:=C([t_{1},t_{1}+h],\mathbb{C}^{d}),
\end{equation*}
their relevant norms
\begin{equation*}
||\cdot||:=\max_{t\in [t_{1}-\tau_{d},t_{1}]}|\cdot|,
\end{equation*}
and
\begin{equation*}
||\cdot||^{+}:=\max_{t\in [t_{1},t_{1}+h]}|\cdot|,
\end{equation*}
and the space
\begin{equation*}
\mathcal{P}^{*}:=C([t_{1}-\tau_{d},t_{1}+h],\mathbb{C}^{d}),
\end{equation*}
with the map $L:\mathcal{P}\times \mathcal{P}^{+}\to \mathcal{P}^{*}$ satisfying
\begin{equation*}
L(\phi_{\delta }, z)(\eta)=
\begin{cases}
\phi_{\delta }(t_{0})+\int_{t_{1}}^{\eta}z(\theta)d \theta, ~\text{if}~ \eta \in [t_{1},t_{1}+h], \\
\phi_{\delta }(\eta), ~~~~~~~~~~~~~~~~~~\text{if} ~\eta \in [t_{1}-\tau_{d},t_{1}].
\end{cases}
\end{equation*}
According to \cite{breda2012approximation}, the map $L$ can be divided into two operators $L_{1}:\mathcal{P}\to  \mathcal{P}^{*}$ and $L_{2}:\mathcal{P}^{+}\to  \mathcal{P}^{*}$ with
\begin{equation}\label{Lproperty1}
L(\phi_{\delta },\omega)=L_{1}\phi_{\delta }+L_{2}\omega,~
\end{equation}
where $(\phi_{\delta },\omega)\in \mathcal{P}\times \mathcal{P}^{+}$, $L_{1}\phi_{\delta }=L(\phi_{\delta },0)$ and $L_{2}\omega=L(0,\omega)$.

In addition, we define the linear operator $\Theta: \mathcal{P}^{*}\to  \mathcal{P}^{+}$ via
\begin{equation}\label{fde}
[\Theta v](t)=F(t,v(t),v_{d}(t)),
\end{equation}
where $v\in \mathcal{P}^{*}$, $t\in [t_{1},t_{1}+h]$ and $v_d(t)=v(t-\tau_d)$.
The fixed point problem
\begin{equation}\label{sub-ope-func}
\omega^{*}=\Theta L(\phi_{\delta },\omega^{*}).
\end{equation}
has a fixed point $\omega^{*}\in  \mathcal{P}^{+}$ if the original problem (\ref{abstact-vari}) has a solution in $[t_{1},t_{1}+h]$. So $\omega^{*}$ satisfies
\begin{equation}\label{ope-func}
U(t_{1}+h,t_{1}) \phi_{\delta }=L(\phi_{\delta },\omega^{*}).
\end{equation}
According to Eq.~(\ref{Lproperty1}), Eq.~(\ref{sub-ope-func}) can be rewritten as
\begin{equation}\label{tran-ope-func}
(I_{\mathcal{P}^{+}}-\Theta L_{2})\omega^{*}=\Theta L_{1}\phi_{\delta }
\end{equation}
where $I_{\mathcal{P}^{+}}$ is the identity operator for the space $\mathcal{P}^{+}$.
Therefore, we can derive the following properties for the operators $\Theta L_{1}$ and $\Theta L_{2}$.
\begin{prop}\label{bound-operator1}
If the operator $\Theta$ is defined by Eq.~(\ref{fde}), it is a bounded linear operator with $v\in \mathcal{P}^{*} $.
\end{prop}

\begin{prop}\label{bound-operator2}
If $L_{1}$ and $L_{2}$ are defined by Eq.~(\ref{Lproperty1}), then $\Theta L_{1}:\mathcal{P}\to\mathcal{P}^+$ and $\Theta L_{2}:\mathcal{P}^+\to\mathcal{P}^+$ are bounded linear operators with regard to $\omega \in \mathcal{P}^{+} $.
\end{prop}

\subsection{Approximation of the evaluation operator}
Since system (\ref{abstact-vari}) can be approximated by large finite ODE systems, the approximated operators are constructed through discretisation by introducing the relevant discrete space of $\mathcal{P}$ and $\mathcal{P}^{+}$ along with the following operators. As large finite ODE systems can be obtained from the modified Euler integration, we can adopt linear interpolation to discretise the space $\mathcal{P}$ and $\mathcal{P}^{+}$.

First of all, based on the time step $h$, consider the mesh $\Lambda _{N+1}:=(t_{1}-Nh,\cdots,t_{1}-h,t_{1})$ in $[t_{1}-\tau_{d},t_{1}]$. We  construct a restriction operator $r_{h}:\mathcal{P} \to \mathcal{P}_{N+1}:=\mathbb{C}^{d(N+1)}$ on $\Lambda _{N+1}$, such that $r_{h}\phi_{\delta }\in \mathcal{P}_{N+1}$, where $[r_{h}\phi_{\delta }]_{i}=\phi_{\delta }(t_{1}-(N+1-i)h) \in \mathbb{C}^{d} $. In addition, there exists a prolongation operator on the mesh $\Lambda _{N+1} $ such that for any $\varpi_{N+1}:=(\varpi^{T}(t_{1}-Nh),\cdots,\varpi^{T}(t_{1}))^{T} \in  \mathcal{P}_{N+1}$, where $\varpi \in \mathcal{P}$, $\bar{r}_{h}:t\in [t_{1}-\tau_{d}, t_{1}] \to \bar{r}_{h}(t)\in \mathbb{C}^{1\times d(N+1)}$,
$
\bar{r}_{h}(t_{1}-(N+1-i)h)\varpi_{N+1}=\varpi(t_{1}-(N+1-i)h),~ i\in \mathbb{Z}[1,N+1],
$
and $\bar{r}_{h}(t)\varpi_{N+1}$ is a polynomial with a degree less than or equal to $2$.

Similarly, consider the mesh $\Lambda _{K+1}:=(t_{1},t_{1}+h_{s},\cdots,t_{1}+Kh_{s})$ in $[t_{1},t_{1}+h]$, where $0<h_{s}<h$, $K=h/ h_{s}$,
the space $\mathcal{P}^{+}$ can be discretised by the restriction operator $R_{h_{s}}: \mathcal{P}^{+} \to \mathcal{P}_{K+1}^{+}:=\mathbb{C}^{d(K+1)}$
on the mesh $\Lambda _{K+1} $ such that $R_{h_s}\psi \in \mathcal{P}_{K+1}^{+}$,
where $R_{h_s}\psi^{i}=\psi (t_{1}+(i-1)h_{s})\in \mathbb{C}^{d}$.
For mesh $\Lambda_{K+1}$ we construct a relevant prolongation operator as follows. For any $\varpi_{K+1}:=(\varpi^{T}(t_{1}),\cdots,\varpi^{T}(t_{1}+Kh_{s})) \in \mathcal{P}_{K+1}^{+}$, where $\varpi \in \mathcal{P}^{+}$,
$\bar{R}_{h_s}: t\in[t_{1},t_{1}+h] \to  \bar{R}_{h_s}(t)\in  \mathbb{C}^{d(K+1)} $, such that
$\bar{R}_{h_{s}}(t_{1}+(i-1)h_{s}) \varpi_{K+1}=\varpi(t_{1}+(i-1)h_{s})$, $i\in \mathbb{Z}[1,K+1]$, and $ \bar{R}_{h_{s}}(t)\varpi_{K+1}$ is a polynomial with degree less than or equal to $K+1$. Here, the operator $\mathfrak{L}:=\bar{R}_{h_{s}}(t)R_{h_{s}}$ is a Lagrange operator \cite{varma1973summability}.

Let $K=1$ ( i.e. $h_{s}=h$ ) and for any given $N$,
the relevant approximated operator $U_{N+1,1}(t_{1}+h,t_{1}):  \mathcal{P}_{N+1}\to  \mathcal{P}_{N+1}$ satisfies
\begin{equation}\label{app-ope}
U_{N+1,2}(t_{1}+h,t_{1})\Phi=r_{h}L(\bar{r}_{h}(t-\tau_{d})\Phi,\bar{R}_{h_{s}}(t)\Psi^{*}),
\end{equation}
where $t\in [t_{1},t_{1}+h]$, $\Phi\in  \mathcal{P}_{N+1}$
and $\Psi^{*}\in \mathcal{P}_{K+1}^{+}$, which is the solution of the following equation
\begin{equation}\label{fix-matrx}
\Psi^{*}=R_{h_{s}}\Theta L(\bar{r}_{h}(t-\tau_{d})\Phi,\bar{R}_{h_{s}}(t)\Psi^{*}).
\end{equation}
It is worth noting that the operator $\bar{R}_{h_{s}}$ at the time interval $[t_{1},t_{1}+h]$ can be more accurate if the time step $h$ is reduced.

\subsection{Convergence analysis for the nonzero eigenvalues of the Jacobian matrix }

In this section, we will present the convergence analysis for $0<\tau_{d}<T$ only. The proof for $\tau_{d}\ge T$ is similar, so will be omitted here.
In order to ensure a unique solution for the initial problem (\ref{abstact-vari}), we introduce the subspace $\mathcal{P}_{Lip}^{+}$ of $\mathcal{P}^{+}$
with the norm
\begin{equation*}
||\psi||_{Lip}^{+}=l(\psi)+||\psi||^{+},~\psi\in \mathcal{P}_{Lip}^{+},
\end{equation*}
where $l(\psi)$ is the Lipschitz constant of $\psi$,
and the subspace $\mathcal{P}_{Lip}$ of $\mathcal{P}$
with the norm as
\begin{equation*}
||\psi||_{Lip}=l(\psi)+||\psi||,~\psi\in \mathcal{P}_{Lip}.
\end{equation*}

To carry out convergence analysis for the eigenvalues of Jacobian of the Poincar\'e map (\ref{poin-map}), the following lemmas are given based on \cite{breda2012approximation}. 

\begin{lemma}\label{L1}
For any $\sigma_{1}^{*}, \sigma_{2}^{*}\in \mathcal{P}^{+}$,
\begin{equation}\label{app-1}
\sigma_{1}^{*}=\mathfrak{L} \Theta L(\phi_{\delta },\sigma_{1}^{*}), ~\phi_{\delta }\in \mathcal{P} ,
\end{equation}
and
\begin{equation*}\label{app-2}
\sigma_{2}^{*}=\Theta L(\phi_{\delta },\sigma_{2}^{*}),~\phi_{\delta }\in \mathcal{P} ,
\end{equation*}
for sufficiently small $h$, and we have
\begin{equation}\label{lem-1}
||\sigma_{1}^{*}-\sigma_{2}^{*}||^{+}\leq c_{1}h^{2},
\end{equation}
where $c_{1}$ is a positive constant.
\end{lemma}

Based on Eq.~(\ref{app-ope}), a new operator in the interval $[t_{1},t_{1}+h]$ can be introduced as
\begin{equation}\label{approximate-1}
\bar{U}_{N+1,2}(t_{1}+h,t_{1})=\bar{r}_{h}U_{N+1,2}(t_{1}+h,t_{1}) r_{h}: \mathcal{P}\to \mathcal{P},
\end{equation}
which has the same geometric and partial multiplicities as the operator $U_{N+1,2}(t_{1}+h,t_{1})$ in Eq.~(\ref{app-ope}). Therefore, there exists a map $\bar{U}_{2}(t_{1}+h,t_{1}):\mathcal{P}\to \mathcal{P}$ such that
\begin{equation}\label{middle-oper}
\bar{U}_{2}(t_{1}+h,t_{1})\phi_{\delta }=L(\phi_{\delta },\sigma^{*}),~  \phi_{\delta }\in \mathcal{P},
\end{equation}
where $\sigma^{*}\in \mathcal{P}^{+}$ is the solution of Eq.~(\ref{app-1}), and $\bar{U}_{N+1,2}(t_{1}+h,t_{1})$ can be written as
\begin{equation*}
\bar{U}_{N+1,2}(t_{1}+h,t_{1})=\mathfrak{L} \bar{U}_{2}(t_{1}+h,t_{1})\mathfrak{L}.
\end{equation*}

\begin{lemma}\label{tiny-map-result}
If the operator $\bar{U}_{2}(t_{1}+h,t_{1})$ is defined as Eq.~(\ref{middle-oper}), we have
\begin{equation}\label{lem-2}
|| \bar{U}_{2}(t_{1}+h,t_{1})-U(t_{1}+h,t_{1})||\leq {\color{red}c_{3}}h^{3},
\end{equation}
where $c_{3}$ is a positive constant.
\end{lemma}

It is worth noting that the evolution operator  $\bar{U}_{2}(t_{1}+ih,t_{1}+(i-1)h)$, where $i=1,\cdots,\bar{N}$ and $\bar{N}:=N+n+\triangle N+1$, must have the same properties as  the operator $U(t_{1}+ih,t_{1}+(i-1)h)$ in the inequality (\ref{lem-2}). Thus, the Poincar\'e map can be obtained by combining all the evolution operators $U(t_{1}+ih,t_{1}+(i-1)h)$ over the entire time interval $[t_{1},t_{1}+T]$. As a result, the convergence problem is equivalent to studying the convergence of the operator $\prod_{i=1}^{\bar{N}}\bar{U}_{2}(t_{1}+ih,t_{1}+(i-1)h)$ to $U(t_{1},t_{1}+T)$.

\begin{lemma}\label{L3}
For the entire interval $[t_{1},t_{1}+T]$ and a sufficiently small time step $h$, we can obtain
\begin{equation}\label{main-th-3}
||U(t_{1}+T, t_{1})- \prod_{i=1}^{\bar{N}}\bar{U}_{2}(t_{1}+ih,t_{1}+(i-1)h)||\leq c_{4}h^{2},
\end{equation}
where $i=1,2,\cdots,\bar{N}$, $\bar{N}:=N+n+\triangle N+1$, and $c_{4}$ is a positive constant.
\end{lemma}

Combining the inequality (\ref{main-th-3}) with the results in \cite{chatelin1973convergence, chatelin2011spectral} and Theorem 4.6 and 4.7 in \cite{breda2012approximation}, the following lemma can be obtained.

\begin{lemma}
 Let $\lambda\in \mathbb{C}\setminus\{0\}$ be an isolated eigenvalue for the operator $U(t_{1}+T,t_{1})$ with the finite algebraic multiplicity $m_{a}$ and ascent (length of longest Jordan chain) $\kappa$, and $\Gamma$ be a neighborhood of $\lambda$ for $\lambda$ of $U$ on the time interval $[t_{1},t_{1}+T]$. For a sufficiently small $h$, $\bar{U}_{2}(t_{1}+T,t_{1})$ has $m$ eigenvalues $\lambda_{2,\iota}$, where $\iota=1,.\dots,m_{a}$, and we have
\begin{equation}\label{lem-3}
\max_{\iota=1,\dots,m_{a}}|\lambda-\lambda_{2,\iota}|\leq c_{5}h^{\frac{2}{\kappa}},
\end{equation}
where $c_{5}$ is a positive constant.
\end{lemma}

It should be noted that $\bar{U}_{N+1,2}$ and $\bar{U}_{2}$ have the same nonzero eigenvalues, geometric and partial multiplicities and eigenvectors. This leads to the following theorem.

\begin{theorem}
 Let $\lambda\in \mathbb{C}\setminus\{0\}$ be an isolated eigenvalue for the operator $U(t_{1}+T,t_{1})$ with the finite algebraic multiplicity $m_{a}$ and the ascent $\kappa$, and let $\Gamma$ be a neighborhood of $\lambda$ for the time interval $[t_{1},t_{1}+T]$. For a sufficiently small $h$, $\bar{U}_{N+1,2}(t_{1}+T,t_{1})$ has $m$ eigenvalues $\lambda_{N+1,2,\iota}$, where $\iota=1,.\dots,m_{a}$ and we have
\begin{equation}\label{Final-resul}
\max_{\iota=1,\dots,m_{a}}|\lambda-\lambda_{N+1,2,\iota}|\leq c_{6}h^{\frac{2}{\kappa}},
\end{equation}
where $c_{6}$ is a positive constant.
\end{theorem}
The inequality (\ref{Final-resul}) holds for any interval $[t_{m},t_{m}+T]$.
From the above study, we can ensure that our proposed approximation method has the expected convergence rate on the nonzero characteristic multipliers of the system (\ref{abstact-vari}).
So our approximation for the Jacobian of the Poincar\'e map (\ref{poin-map}) is reliable.
It is also worth noting that by adopting a high-order integration method (e.g. Runge-Kutta method) with a sufficiently small time step $h$, the approximated operator could be more accurate $O(h^{4})$. However, this would also require higher-order corrections at the crossing and grazing events for the terms derived in Section \ref{modification-discontinuous}. Without these corrections
the convergence of the approximated operator cannot be guaranteed as the same with the order of the numerical integration. Furthermore, if the system encounters sufficiently large number  grazing events, the convergence rate will be lower than $O(h^{2})$ due to these grazing events.

\section{Calculation of the Lyapunov exponents}\label{sec-computation}
The dynamics of system (\ref{eq-hybsys-main}) can be represented by the Poincar\'e map (\ref{poin-map}) as
\begin{equation}\label{wholepoincare}
Y_{m+1,0}=P^{m}(Y_{1,0})=P\circ \cdots P\circ P(Y_{1,0}) ,
\end{equation}
where the Jaobian matrix of $P^{m}$ is $\prod^{m}_{i=1}M_{i}$.
According to Definition \ref{LEdis},
LEs can be calculated as
\begin{equation}\label{originLE}
\vartheta_{i}=\lim_{m\to \infty }\frac{1}{m}\ln|\lambda^{m}_{i}|,~i=1,\cdots,d(N+1),
\end{equation}
where $\lambda^{m}_{i} $ is the $i^\text{th}$ eigenvalues of $\prod^{m}_{i=1}M_{i}$ .

However, calculating LEs by using Eq.~(\ref{originLE}) will introduce an overflow problem.
Specifically, some elements of the Jacobian matrix will be very large for chaotic attractors, and some of them could be very small for periodic attractors, which may cause inaccuracies.
On the other hand, calculating LEs from the Jacobian matrix directly is time-consuming as the time-delayed dynamical system is high-dimensional. To overcome these issues , LEs can be computed according to the average exponential divergence rate between the basis orbit started from $Y_1(0)$ and its neighborhood orbit along the direction of $v_{1,0}=\frac{Y_{1,0}}{||Y_{1,0}||}$ as
\begin{equation}\label{modiLE}
\vartheta(Y_{1,0},v_{1,0})=\lim_{m\to \infty }\frac{1}{m}\ln \frac{||\delta Y_{m,0}||}{||\delta Y_{1,0}||},
\end{equation}
where $||\delta Y_{m,0}||$ is the norm of $\delta Y_{m,0}$ and $m\in \mathbb{Z}^{+}$.

Next, choose $Y_{1,0}\in  \mathbb{R}^{d(N+1)}$, and its related linearly independent initial perturbed vector $(\delta Y^{1}_{1,0},\delta Y^{2}_{1,0},$ $\cdots,\delta Y^{d(N+1)}_{1,0})$ can be normalised as
\begin{align}\label{normalise}
(\delta v^{1}_{1,0},\delta v^{2}_{1,0},\cdots,\delta v^{d(N+1)}_{1,0})=(\frac{\delta Y^{1}_{1,0}}{||\delta Y^{1}_{1,0}||},\frac{\delta Y^{2}_{1,0}}{||\delta Y^{2}_{1,0}||},\cdots,\frac{\delta Y^{d(N+1)}_{1,0}}{||\delta Y^{d(N+1)}_{1,0}||}).
\end{align}
Substituting the vector (\ref{normalise}) to Eq.~(\ref{wholepoincare}) obtains the second vector $(\delta Y^{1}_{2,0},\delta Y^{2}_{2,0},\cdots,\delta Y^{d(N+1)}_{2,0})$, and Gram-Schmidt orthonormalization \cite{ stoer2013introduction} can be applied to normalise the second vector, which gives a new vector $(\delta v^{1}_{2,0},\delta v^{2}_{2,0},\cdots,\delta v^{d(N+1)}_{2,0})$.
For the next iteration, the second vector will be used as the initial vector to be substituted into Eq.~(\ref{wholepoincare}). Likewise, repeating $m$ times for this process gives the $m^\text{th}$ vector $(\delta Y^{1}_{m,0},\delta Y^{2}_{m,0},\cdots,\delta Y^{d(N+1)}_{m,0})$.
The steps of Gram-Schmidt orthonormalization are given as follows
\begin{align*}
~~~~~~~~~~~~~~~~~~~~~~~~~~~~~V^{1}_{m,0}=&\,\delta Y^{1}_{m,0},\\
\delta v^{1}_{m,0}=&\,\frac{V^{1}_{m,0}}{||V^{1}_{m,0}||} ,\\
V^{2}_{m,0}=&\,\delta Y^{2}_{m,0}-<\delta Y^{2}_{m,0},\delta v^{1}_{m,0}>\delta v^{1}_{m,0},\\
\delta v^{2}_{m,0}=&\,\frac{V^{2}_{K}(0)}{||V^{2}_{m,0}||},\\
&\vdots\\
V^{d(N+1)}_{m,0}=&\,\delta Y^{2(N+1)}_{m,0}-<\delta Y^{2(N+1)}_{m,0},\delta v^{1}_{m,0}>\delta v^{1}_{m,0}-\cdots\\
&\,-<\delta Y^{d(N+1)}_{m,0},\delta v^{d(N+1)-1}_{m,0}>\delta v^{d(N+1)-1}_{m,0},
\end{align*}
\begin{align*}
\delta v_{m,0}^{d(N+1)}=\,\frac{V^{d(N+1)}_{m,0}}{||V^{d(N+1)}_{m,0}||} ,~~~~~~~~~~~~~
\end{align*}
where $||V^{i}_{m,0}||$ is the norm of $V^{i}_{m,0}$,     $\langle \delta Y^{i}_{m,0} , \delta v^{\bar{i}}_{m,0} \rangle$  
 $(i,\bar{i}=1,2,\cdots,d(N+1))$ is a standard scalar product.
Finally, LEs can be calculated by using
\begin{align}\label{LEfor}
\vartheta_{i}\approx\frac{1}{m}\ln \prod^{m}_{\varrho=1}||V^{i}_{\varrho}(0)||=\frac{1}{m}\sum^{m}_{\varrho=1}\ln ||V^{i}_{\varrho}(0)||.
\end{align}

\begin{remark}
Based on the above analysis, a guideline for the implementation of the algorithm is presented as follows.

\textbf {Step 1}: Calculate the Jacobian matrix according to the relevant trajectory at the time step after the system is stabilised by the time-delayed feedback controller;

\textbf {Step 2}: If the trajectory approaches to grazing, calculate its relevant Jacobian using Eq.~(\ref{matrixfuncs-ca1}) or Eq.~(\ref{matrixfuncs-ca2}), and then insert it to the matrix $M_m$ in Eq.~(\ref{mainvar}) at the grazing moment;

\textbf {Step 3}: Choose appropriate initial perturbed unit vectors, and calculate the Floquet Multipliers of each Poincar\'e map using Gram-Schmidt orthonormalization;

\textbf {Step 4}: Calculate the LEs using Eq.~(\ref{LEfor}) after several evolutions of Poincar\'e map.

\end{remark}

\section{Numerical studies}\label{sec-simulation}

In this section, we will show the effectiveness of our proposed method by studying the soft impacting system with a delayed feedback controller presented in Fig.~\ref{fig-model}. Since the system has many coexisting attractors when grazing is encountered \cite{liu2017controlling}, our control objective here is to drive the system from its current attractor to a desired one. Calculating the LEs of the system allows us to monitor the stability of the delayed feedback controller and its effective parametric regime.

\begin{figure}[h!]
\centering
\includegraphics[width=0.55\textwidth]{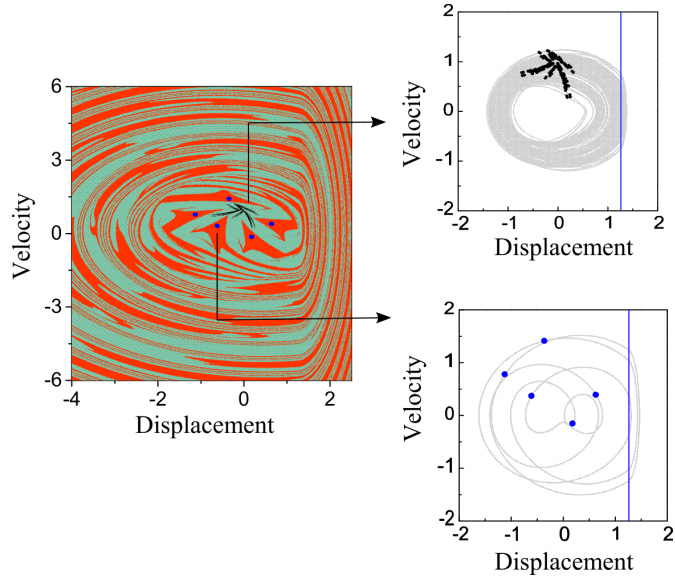}
\caption{Basin of attraction of the impacting system computed for $\zeta= 0.01, ~e = 1.26, ~a = 0.7, ~\beta = 28 ~ \textrm{and}~ \omega= 0.802$.
Black dots denote the chaotic attractor with green basin, blue dots represent the period-$5$ attractor with red basin, and blue lines denote the impact boundary.}\label{coexist}
\end{figure}

We choose the following parameters for the impacting system,
\begin{equation*}\label{parameter}
\zeta= 0.01, ~e = 1.26, ~a = 0.7, ~\beta = 28 ~ \textrm{and}~ \omega= 0.802.
\end{equation*}
For these parameters a grazing event is encountered, and a chaotic and a period-$5$ attractors coexist as shown in Fig.~\ref{coexist}.

\subsection{Case $\tau_{d} \ge T$}
\begin{figure}[h!]
\centering
\includegraphics[width=0.6\textwidth]{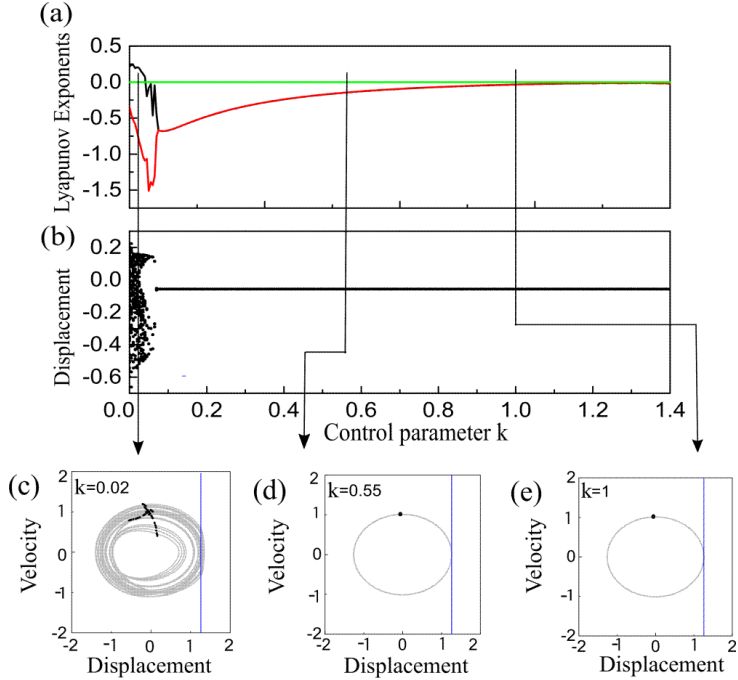}
\caption{(a) LEs and (b) displacement of the impacting system under the delayed feedback controller as functions of the control parameter $k$. Black, red and green lines denote the two largest LEs and the zero line, respectively. Additional panels show the phase trajectories of the system calculated for (c) $k=0.02$, (d) $k=0.55$ and (e) $k=1$. Black dots represent the Poincar\'e sections, and blue lines represent the impact boundary.
}\label{LElarge}
\end{figure}

Fig.~\ref{LElarge} presents the first example of using the delayed feedback controller (\ref{eq-dde-u}) for which a large delayed time (i.e. $\tau_{d} \ge T$) was considered, and the control parameter $k$ was varied from $0$ to $1.4$. As can be seen from Fig.~\ref{LElarge}(a), the largest LEs are all greater than $0$ for $k\in [0,0.04]$ and the system presents a chaotic motion as shown in Fig.~\ref{LElarge}(b). The phase trajectory of the chaotic motion for $k=0.02$ is presented in Fig.~\ref{LElarge}(c). For $k\in(0.04,0.055)$, the largest LEs decrease and suddenly increase to the neighbourhood of zero at $k=0.055$ indicating a period doubling of the system. Similarly, at $k=0.065$, such a fluctuation is observed again. Thereafter, the largest LEs decrease dramatically, and then increase gradually from $k=0.07$. For $k\in[0.07,1.4]$, both LEs are below zero, and the system has period-$1$ response which is demonstrated by Figs.~\ref{LElarge}(d) and (e).

\begin{figure}[h!]
\centering
\includegraphics[width=0.75\textwidth]{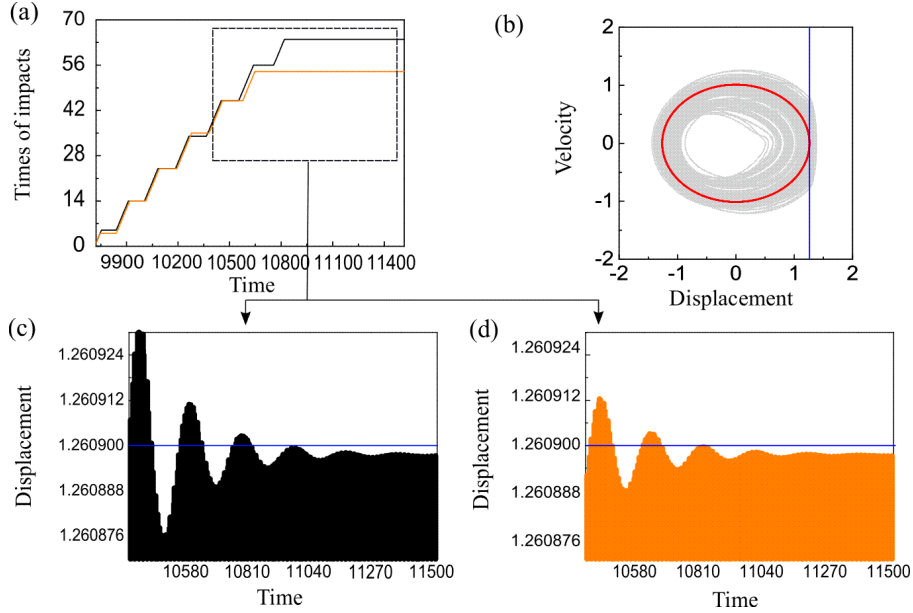}
\caption{(a) Number of impacts as a function of time without (black line) and with (orange line) the grazing estimation algorithm based on the discontinuous condition calculated for $\zeta= 0.01$, $e = 1.2609$, $a = 0.7$, $\beta = 28$, $ \omega= 0.802$ and $k=1.4$. (b) Phase trajectories of the impacting system controlled from chaotic (grey line) to period-1 (red line) response. Time histories of displacement of the system (c) without and (d) with the algorithm are presented, and blue lines indicate the discontinuous boundary.
}\label{comparison}
\end{figure}

A critical issue for computing nonsmooth dynamical systems is that the accumulated computational error from the impact boundary due to grazing event could lead to inaccurate simulation. Fig.~\ref{comparison} compares the computations of the impacting system for $e=1.2609$ controlled from a chaotic response to a period-$1$ response by using the delayed feedback control with and without the grazing estimation algorithm. The number of impacts as a function of time without (black line) and with (orange line) the grazing estimation algorithm is presented in Fig.~\ref{comparison}(a) which were counted from $t=9722$, and the phase trajectories from chaotic (grey line) to period-$1$ (red line) response are shown in Fig.~\ref{comparison}(b). It can be seen from the figure that the accumulated error was built up in the number of impacts, and a clear difference can be observed from $t=10411$. The cause of such a difference can be found from Figs.~\ref{comparison}(c) and (d), where the time histories of displacement of the impacting system are shown. As can be seen from the figures, the system with the grazing estimation algorithm was stabilised quicker than the one without the algorithm. 

\subsection{Case $0<\tau_{d}<T$}
\begin{figure}[h!]
\centering
\includegraphics[width=0.85\textwidth]{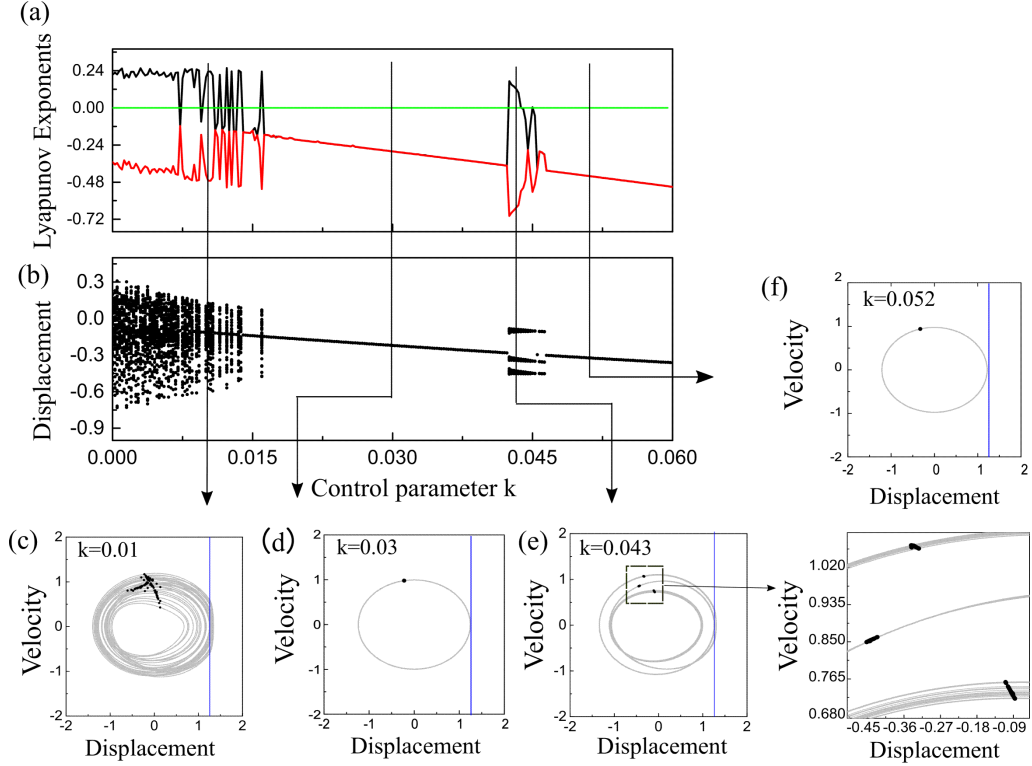}
\caption{(a) LEs and (b) displacement of the impacting system under the delayed feedback controller as functions of the control parameter $k$. Black, red and green lines denote the two largest LEs and the zero line, respectively. Phase trajectories of the system calculated for (c) $k=0.01$, (d) $k=0.03$, (e) $k=0.043$ and (f) $k=0.052$ are shown. Black dots represent the Poincar\'e sections, and blue lines indicate the nonsmooth boundary.
}\label{LEtiny}
\end{figure}

For the case of a small time delay (i.e. $0<\tau_{d}<T$), we present the example for $\tau_{d}=T/2$ in Fig.~\ref{LEtiny}. It can be seen from the figures that the system has chaotic motion for $k\in [0,0.007]$ and its largest LEs are all greater than zero (green line). For $k\in (0.007,0.015]$, the system experiences transient periodic motion, and the relevant largest LEs are smaller than zero which is consistent with the result shown in Fig.~\ref{LEtiny}(b) indicating several alternations between chaotic and periodic motions. At $k=0.016$, the system has a very narrow chaotic window and bifurcates into a non-impact period-$1$ response immediately lasting until $k=0.0425$ at where another chaotic regime is encountered. For $k\in [0.0425,0.045]$, the system has chaotic response in most of the region, but has a small window of period-$3$ response in $k\in[0.044,0.04475]$. After $k=0.045$, the non-impact period-$1$ response emerges again as the control parameter $k$ increases. To compare Figs.~\ref{LEtiny}(a) and (b), the evolution of the calculated LEs is consistent with system's bifurcation, which is also demonstrated by the phase trajectories presented in Figs.~\ref{LEtiny}(c)-(f).

\section{Conclusions}\label{sec-conclusion}
This paper studies a numerical method for calculating the LEs of time-delayed piecewise-smooth systems by using a soft impacting system under the delayed feedback control with a particular focus on its near-grazing dynamics. The main feature of the proposed method is that it can provide improved accuracy for the stability analysis of periodic orbits by estimating the point of discontinuity locally along trajectories of piecewise-smooth DDEs with an accuracy of the same order as its integration method. In addition, the method can also be applied to the other nonsmooth dynamical systems with a delayed argument, such that it can be used as a generic computational tool for stability analysis.

The main tasks were to build an effective variational equation and obtain the Jacobian for the delayed impacting system. As the delayed impacting system is infinite dimensional, it was approximated by finite dimensional systems, which were discretised by the modified Euler integration method at each time step. Then the DDE system converted to a time-discrete map by constructing a Poincar\'e map, and its linearisation was introduced to obtain its variational equation. Then the Jacobian of the map was obtained by combining all the approximating systems linearised from the variational equation at each time step in one period of external excitation. In order to increase the convergence rate and improve computational accuracy, a grazing estimation algorithm was introduced. The convergence rate of eigenvalues of the Jacobian matrix was studied by using the spectral theory of the evolutionary operator. In particular, the delayed impacting system was described as an evolutionary operator with the expected convergence rate for the relevant nonzero eigenvalues of the Jacobian, therefore guaranteeing the reliability of the proposed numerical method.

Our numerical studies considered two scenarios of delay time in the system, a larger ($\tau_{d}\ge T$) and a smaller ($0<\tau_{d}<T$) delay than the period of excitation. Both cases showed that the calculated LEs were consistent with the bifurcation of the system, and the grazing estimation algorithm had improved accuracy for simulating nonsmooth dynamical systems.

\section*{Acknowledgements}
This work has been supported by EPSRC under Grant No. EP/P023983/1. Mr Zhi Zhang would like to acknowledge the financial support from the University of Exeter for his Exeter International Excellence Scholarship.
JS' research is  supported by EPSRC Fellowship EP/N023544/1 and the European Union’s Horizon 2020 research and innovation programme under grant agreement number 820970, project TiPES.

\section*{Appendix.}

\textit{Proof of Proposition \ref{bound-operator1}}:
Let $v_{1},v_{2} \in \mathcal{P}^{*} $, where
\begin{equation*}
\Theta v_{1}(t)=F(t, v_{1}(t),v_{1}(t-\tau_{d})),
\end{equation*}
and
\begin{equation*}
\Theta v_{2}(t)=F(t, v_{2}(t),v_{2}(t-\tau_{d})).
\end{equation*}
Then we can obtain
\begin{align*}
||\Theta (v_{1} +v_{2})||\leq&\, |F_{\bar{j},1}(t)v_{1}(t)+F_{\bar{j},2}(t)v_{1}(t-\tau_{d}) |+| F_{j,1}(t)v_{2}(t)+F_{j,2}(t)v_{2}(t-\tau_{d})| \\
=&\,|| \Theta v_{1}||+|| \Theta v_{2}||
\end{align*}
In addition, according to Eq.~(\ref{linear-property}), there must exist a positive constant $B_{\Theta}$ satisfying that, for any $ v\in  \mathcal{P}^{*}  $, $|| \Theta v||\leq B_{\Theta} || v||$. Therefore, the operator $\Theta$ is bounded and linear in the space.

\vspace{10pt}

\textit{Proof of Proposition \ref{bound-operator2}}:
For $\forall \phi_{\delta }$, there exist $\omega_{0}, \omega_{1}, \omega_{2}\in \mathcal{P}^{+}$ such that $\Theta L(\phi_{\delta },\omega_{0})=\omega_{0}$ and $\omega_{0}=\omega_{1}+\omega_{2}$.
So we have
\begin{align*}
\Theta L(\phi_{\delta },\omega_{0})=&\,\Theta [L_{1}(\phi_{\delta },\omega_{1})+L_{2}\omega_{2}]\\
=&\,\Theta L_{1}(\phi_{\delta })+\Theta L_{2}\omega_{1}+\Theta L_{2}\omega_{2}\\
=&\,\Theta L_{1}(\phi_{\delta })+\Theta L_{2}(\omega_{1}+\omega_{2}).
\end{align*}
According to the Eqs.~(\ref{fde})-(\ref{ope-func}), if
$\Theta L(0,\omega)=\omega$ (where $\omega\in \mathcal{P}^{+}$) holds,
$L(0,\omega)$ must be the solution of the following system
\begin{equation}\label{L-2-origi}
\begin{cases}
\frac{\mathrm{d} }{\mathrm{d}t}\delta u_{0}(t)=F(t,\delta u_{0}(t),\delta u_{d}(t)),\\
\delta u_{0}(s)=0,
\end{cases}
\end{equation}
where $F\in C(\mathbb{P},\mathbb{R}^{d})$ and $s\in [t_{1}-\tau_{d},t_{1}]$.
Then for any $ \omega\in \mathcal{P}^{+}$, it gives $|| \Theta L_{2}\omega||=||\omega||$, so $\Theta L_{2}$ is bounded.

Let $\phi_{\delta,1}, \phi_{\delta,2}\in \mathcal{P}$,
and for $\phi_{\delta,1}+\phi_{\delta,2}$, there exists $\omega\in  \mathcal{P}^{+}$ such that $\Theta L(\phi_{\delta,1}+\phi_{\delta,2},\omega)=\omega$.
Also, there exists $\omega_{1},~\omega_{2}\in \mathcal{P}^{+}$, such that $\omega=\omega_{1}+\omega_{2}$.
Then we have
\begin{align*}
 L(\phi_{\delta,1}+\phi_{\delta,2},\omega)=&\,L(\phi_{\delta,1},\omega_{1})+L(\phi_{\delta,2},\omega_{2})\\
 =&\,L_{1}\phi_{\delta,1}+L_{2}\omega_{1}+L_{1}\phi_{\delta,2}+L_{2}\omega_{2}\\
 =&\,L_{1}(\phi_{\delta,1}+\phi_{\delta,2})+L_{2}\omega
\end{align*}
Since
\begin{align*}
\Theta [L_{1}\phi_{\delta,1}+L_{2}\omega_{1}+L_{1}\phi_{\delta,2}+L_{2}\omega_{2}]
=\Theta L_{1}\phi_{\delta,1}+\Theta L_{1}\phi_{\delta,2}+\Theta L_{2}\omega
\end{align*}
and
\begin{equation*}
\Theta[L_{1}(\phi_{\delta,1}+\phi_{\delta,2})+L_{2}\omega]=\Theta L_{1}(\phi_{\delta,1}+\phi_{\delta,2})+\Theta L_{2}\omega,
\end{equation*}
 $\Theta L_{1}$ is a bounded linear operator.

\vspace{10pt}

\textit{Proof of Lemma \ref{L1}}:
Based on Theorem 3.3 in \cite{breda2012approximation} and let $\sigma_{1}^{*}=\sigma_{2}^{*}+\rho^{*}$, we have
\begin{align}\label{lem1-1}
||\rho^{*}||
:=&||\sigma_{1}^{*}-\sigma_{2}^{*}||^{+}=||(I_{\mathcal{P}^{+}}-\mathfrak{L} \Theta L_{2})^{-1} ||~|| (I_{\mathcal{P}^{+}}- \mathfrak{L} )||^{+}||\sigma_{2}^{*}||_{Lip}^{+},
\end{align}
For sufficiently small $h$, $||(I_{\mathcal{P}^{+}}- \mathfrak{L})||^{+}$ is the global error from the modified Euler integration, which satisfies
\begin{align*}
||(I_{\mathcal{P}^{+}}- \mathfrak{L} )||^{+}\leq c_{2}h^{2},
\end{align*}
where $c_{2}$ is a positive constant.
Since
\begin{align*}
I_{\mathcal{P}^{+}}-\mathfrak{L} \Theta L_{2} =(I_{\mathcal{P}^{+}}-\Theta L_{2})+ (I_{\mathcal{P}^{+}}- \mathfrak{L})\Theta L_{2},
\end{align*}
and $\Theta L_{2}$ is bounded, if $h\to 0$,
$(I_{\mathcal{P}^{+}}-\mathfrak{L} \Theta L_{2} )^{-1}=(I_{\mathcal{P}^{+}}-\Theta L_{2}).$
In addition, as
\begin{equation}\label{bound-sigma-1}
\sigma_{2}^{*}=(I_{\mathcal{P}_{Lip}^{+}}-\Theta L_{2})^{-1}\Theta L_{1}\phi_{\delta },
\end{equation}
and
\begin{equation}\label{bound-sigma-2}
||\sigma_{2}^{*}||_{Lip}^{+}\leq || (I_{\mathcal{P}_{Lip}^{+}}-\Theta L_{2})^{-1}||~|| \Theta L_{1}||~|| \phi_{\delta }||_{Lip},
\end{equation}
$|| \sigma_{2}^{*}||^+_{Lip}$ is bounded. Thus, there must exist a positive constant $c_{1}$ for Eq.~(\ref{lem1-1}) satisfying
\begin{equation*}
||\rho^{*}||\leq c_{1}h^{2}.
\end{equation*}

\vspace{10pt}

\textit{Proof of Lemma \ref{tiny-map-result}}:
For $(\phi_{\delta },\omega_{1}^{*}),~ (\phi_{\delta },\omega_{2}^{*}) \in \mathcal{P}_{Lip}^{+}\times \mathcal{P}^{+}$, based on Eq.~(\ref{Lproperty1}), we have
\begin{align*}
|| \bar{U}_{2}(t_{1}+h,t_{1})-U(t_{1}+h,t_{1})||=|| L(\phi_{\delta },\omega_{1}^{*})-L(\phi_{\delta },\omega_{2}^{*})||
=||L_{2}(\omega_{1}^{*}-\omega_{2}^{*})||,
\end{align*}
where
\begin{equation*}
\omega^{*}=\bar{I}_{\mathcal{P}^{+}} \Theta L(\phi_{\delta },\omega_{1}^{*})
\end{equation*}
and
\begin{equation*}
\sigma^{*}=\Theta L(\phi_{\delta },\omega_{2}^{*}).
\end{equation*}
So
\begin{align*}
|| \bar{U}_{2}(t_{1}+h,t_{1})-U(t_{1}+h,t_{1})||=&||L_{2}(\omega_{1}^{*}-\omega_{2}^{*})||\\
=&|| \int^{t_{1}+h}_{t_{1}}(\omega_{1}^{*}-\omega_{2}^{*})(t)\mathrm{d} t||=||(\omega_{1}^{*}-\omega_{2}^{*})||^{+}h.
\end{align*}
According to Eqs.~(\ref{lem1-1}) and (\ref{bound-sigma-1}) and the inequality (\ref{bound-sigma-2}),  we have
\begin{equation*}
|| \bar{U}_{2}(t_{1}+h,t_{1})-U(t_{1}+h,t_{1})||\leq c_{3}h^{3}.
\end{equation*}

\vspace{10pt}

\textit{Proof of Lemma \ref{L3}}:
According to Lemma \ref{tiny-map-result}, we assume that there are two positive constants $M_{1}$ and $M_{2}$ such that
\begin{equation}\label{4-main-1}
||\prod_{i=2}^{\bar{N}-1}U(t_{1}+ih,t_{1}+(i-1)h)||\leq M_{1},
\end{equation}
and
\begin{equation}\label{4-main-2}
||\prod_{j=1}^{\bar{N}-1}\bar{U}_{2}(t_{1}+jh,t_{1}+(j-1)h)||\leq M_{2}.
\end{equation}
Therefore,
\begin{align*}
&||U(t_{1}+T, t_{1})- \prod_{i=1}^{\bar{N}}\bar{U}_{2}(t_{1}+ih,t_{1}+(i-1)h)||\\
\leq& \bar{N}  \prod_{i=2}^{\bar{N}-1}U(t_{1}+ih,t_{1}+(i-1)h) \prod_{j=1}^{\bar{N}-1}\bar{U}_{2}(t_{1}+jh,t_{1}+(j-1)h)   c_{3}h^{3}\\
\leq&\bar{N} M_{1}M_{2}c_{3}h^{3}=\frac{T}{h} M_{1}M_{2}c_{3}h^{3} =c_{4}h^{2}.
\end{align*}

\section*{Compliance with ethical standards}

\section*{Conflict of interest}

The authors declare that they have no conflict of interest concerning the publication of this manuscript.

\section*{Data accessibility}
The datasets generated and analysed during the current study are available from the corresponding author on reasonable request.

\bibliography{bibl}

\providecommand{\noopsort}[1]{}\providecommand{\singleletter}[1]{#1}%
\begin{thebibliography}{53}
\providecommand{\natexlab}[1]{#1}
\providecommand{\url}[1]{\texttt{#1}}
\providecommand{\href}[2]{#2}
\providecommand{\path}[1]{#1}
\providecommand{\eprint}[1]{\href{http://arxiv.org/abs/#1}{\path{#1}}}
\providecommand{\DOIprefix}{doi:}
\providecommand{\ArXivprefix}{arXiv:}
\providecommand{\URLprefix}{URL: }
\providecommand{\Pubmedprefix}{pmid:}
\providecommand{\doi}[1]{\href{http://dx.doi.org/#1}{\path{#1}}}
\providecommand{\Pubmed}[1]{\href{pmid:#1}{\path{#1}}}
\providecommand{\BIBand}{and}
\providecommand{\bibinfo}[2]{#2}
\ifx\xfnm\undefined \def\xfnm[#1]{\unskip,\space#1}\fi
\bibitem[{Bernardo et~al.(2008)Bernardo, Budd, Champneys and
  Kowalczyk}]{bernardo2008piecewise}
\bibinfo{author}{Bernardo\xfnm[ M.]}, \bibinfo{author}{Budd\xfnm[ C.]},
  \bibinfo{author}{Champneys\xfnm[ A.R.]}, \bibinfo{author}{Kowalczyk\xfnm[
  P.]}.
\newblock \bibinfo{title}{Piecewise-smooth dynamical systems: theory and
  applications}; vol. \bibinfo{volume}{163}.
\newblock \bibinfo{publisher}{Springer Science \& Business Media};
  \bibinfo{year}{2008}.
\bibitem[{Thompson and Ghaffari(1982)}]{thompson1982chaos}
\bibinfo{author}{Thompson\xfnm[ J.]}, \bibinfo{author}{Ghaffari\xfnm[ R.]}.
\newblock \bibinfo{title}{Chaos after period-doubling bifurcations in the
  resonance of an impact oscillator}.
\newblock \bibinfo{journal}{Physics Letters A}
  \bibinfo{year}{1982};\bibinfo{volume}{91}(\bibinfo{number}{1}):\bibinfo{pages}{5--8}.
\bibitem[{Muszynska and Goldman(1995)}]{muszynska1995chaotic}
\bibinfo{author}{Muszynska\xfnm[ A.]}, \bibinfo{author}{Goldman\xfnm[ P.]}.
\newblock \bibinfo{title}{Chaotic responses of unbalanced rotor/bearing/stator
  systems with looseness or rubs}.
\newblock \bibinfo{journal}{Chaos, Solitons \& Fractals}
  \bibinfo{year}{1995};\bibinfo{volume}{5}(\bibinfo{number}{9}):\bibinfo{pages}{1683--1704}.
\bibitem[{Yin et~al.(2019)Yin, Ji and Wen}]{yin2019complex}
\bibinfo{author}{Yin\xfnm[ S.]}, \bibinfo{author}{Ji\xfnm[ J.]},
  \bibinfo{author}{Wen\xfnm[ G.]}.
\newblock \bibinfo{title}{Complex near-grazing dynamics in impact oscillators}.
\newblock \bibinfo{journal}{International Journal of Mechanical Sciences}
  \bibinfo{year}{2019};\bibinfo{volume}{156}:\bibinfo{pages}{106--122}.
\bibitem[{Ing et~al.(2010)Ing, Pavlovskaia, Wiercigroch and
  Banerjee}]{ing2010bifurcation}
\bibinfo{author}{Ing\xfnm[ J.]}, \bibinfo{author}{Pavlovskaia\xfnm[ E.]},
  \bibinfo{author}{Wiercigroch\xfnm[ M.]}, \bibinfo{author}{Banerjee\xfnm[
  S.]}.
\newblock \bibinfo{title}{Bifurcation analysis of an impact oscillator with a
  one-sided elastic constraint near grazing}.
\newblock \bibinfo{journal}{Physica D: Nonlinear Phenomena}
  \bibinfo{year}{2010};\bibinfo{volume}{239}(\bibinfo{number}{6}):\bibinfo{pages}{312--321}.
\bibitem[{Jeffrey et~al.(2010)Jeffrey, Champneys, di~Bernardo and
  Shaw}]{jeffrey2010catastrophic}
\bibinfo{author}{Jeffrey\xfnm[ M.R.]}, \bibinfo{author}{Champneys\xfnm[ A.]},
  \bibinfo{author}{di~Bernardo\xfnm[ M.]}, \bibinfo{author}{Shaw\xfnm[ S.]}.
\newblock \bibinfo{title}{Catastrophic sliding bifurcations and onset of
  oscillations in a superconducting resonator}.
\newblock \bibinfo{journal}{Physical Review E}
  \bibinfo{year}{2010};\bibinfo{volume}{81}(\bibinfo{number}{1}):\bibinfo{pages}{016213}.
\bibitem[{Qiu et~al.(2019{\natexlab{a}})Qiu, Sun, Wang and
  Gao}]{qiu2019observer}
\bibinfo{author}{Qiu\xfnm[ J.]}, \bibinfo{author}{Sun\xfnm[ K.]},
  \bibinfo{author}{Wang\xfnm[ T.]}, \bibinfo{author}{Gao\xfnm[ H.]}.
\newblock \bibinfo{title}{Observer-based fuzzy adaptive event-triggered control
  for pure-feedback nonlinear systems with prescribed performance}.
\newblock \bibinfo{journal}{IEEE Transactions on Fuzzy Systems}
  \bibinfo{year}{2019}{\natexlab{a}};\bibinfo{volume}{27}(\bibinfo{number}{11}):\bibinfo{pages}{2152--2162}.
\bibitem[{Ing et~al.(2007)Ing, Pavlovskaia, Wiercigroch and
  Banerjee}]{ing2007experimental}
\bibinfo{author}{Ing\xfnm[ J.]}, \bibinfo{author}{Pavlovskaia\xfnm[ E.]},
  \bibinfo{author}{Wiercigroch\xfnm[ M.]}, \bibinfo{author}{Banerjee\xfnm[
  S.]}.
\newblock \bibinfo{title}{Experimental study of impact oscillator with
  one-sided elastic constraint}.
\newblock \bibinfo{journal}{Phil Trans R Soc A}
  \bibinfo{year}{2007};\bibinfo{volume}{366}(\bibinfo{number}{1866}):\bibinfo{pages}{679--705}.
\bibitem[{Nordmark(1997)}]{nordmark1997universal}
\bibinfo{author}{Nordmark\xfnm[ A.B.]}.
\newblock \bibinfo{title}{Universal limit mapping in grazing bifurcations}.
\newblock \bibinfo{journal}{Physical Review E}
  \bibinfo{year}{1997};\bibinfo{volume}{55}(\bibinfo{number}{1}):\bibinfo{pages}{266}.
\bibitem[{Nordmark(1991)}]{nordmark1991non}
\bibinfo{author}{Nordmark\xfnm[ A.B.]}.
\newblock \bibinfo{title}{Non-periodic motion caused by grazing incidence in an
  impact oscillator}.
\newblock \bibinfo{journal}{Journal of Sound and Vibration}
  \bibinfo{year}{1991};\bibinfo{volume}{145}(\bibinfo{number}{2}):\bibinfo{pages}{279--297}.
\bibitem[{St{\'e}p{\'a}n and Insperger(2006)}]{stepan2006stability}
\bibinfo{author}{St{\'e}p{\'a}n\xfnm[ G.]}, \bibinfo{author}{Insperger\xfnm[
  T.]}.
\newblock \bibinfo{title}{Stability of time-periodic and delayed systems-a
  route to act-and-wait control}.
\newblock \bibinfo{journal}{Annual Reviews in Control}
  \bibinfo{year}{2006};\bibinfo{volume}{30}(\bibinfo{number}{2}):\bibinfo{pages}{159--168}.
\bibitem[{Beregi et~al.(2019)Beregi, Takacs and
  St{\'e}p{\'a}n}]{beregi2019bifurcation}
\bibinfo{author}{Beregi\xfnm[ S.]}, \bibinfo{author}{Takacs\xfnm[ D.]},
  \bibinfo{author}{St{\'e}p{\'a}n\xfnm[ G.]}.
\newblock \bibinfo{title}{Bifurcation analysis of wheel shimmy with non-smooth
  effects and time delay in the tyre--ground contact}.
\newblock \bibinfo{journal}{Nonlinear Dynamics}
  \bibinfo{year}{2019};\bibinfo{volume}{98}(\bibinfo{number}{1}):\bibinfo{pages}{841--858}.
\bibitem[{Zhang et~al.(2011)Zhang, Meng and Song}]{zhang2011dynamics}
\bibinfo{author}{Zhang\xfnm[ T.]}, \bibinfo{author}{Meng\xfnm[ X.]},
  \bibinfo{author}{Song\xfnm[ Y.]}.
\newblock \bibinfo{title}{The dynamics of a high-dimensional delayed pest
  management model with impulsive pesticide input and harvesting prey at
  different fixed moments}.
\newblock \bibinfo{journal}{Nonlinear Dynamics}
  \bibinfo{year}{2011};\bibinfo{volume}{64}(\bibinfo{number}{1-2}):\bibinfo{pages}{1--12}.
\bibitem[{Carvalho and Pinto(2018)}]{carvalho2018new}
\bibinfo{author}{Carvalho\xfnm[ A.R.]}, \bibinfo{author}{Pinto\xfnm[ C.M.]}.
\newblock \bibinfo{title}{New developments on aids-related cancers: The role of
  the delay and treatment options}.
\newblock \bibinfo{journal}{Mathematical Methods in the Applied Sciences}
  \bibinfo{year}{2018};\bibinfo{volume}{41}(\bibinfo{number}{18}):\bibinfo{pages}{8915--8928}.
\bibitem[{Yan et~al.(2017)Yan, Xu and Wiercigroch}]{yan2017basins}
\bibinfo{author}{Yan\xfnm[ Y.]}, \bibinfo{author}{Xu\xfnm[ J.]},
  \bibinfo{author}{Wiercigroch\xfnm[ M.]}.
\newblock \bibinfo{title}{Basins of attraction of the bistable region of
  time-delayed cutting dynamics}.
\newblock \bibinfo{journal}{Physical Review E}
  \bibinfo{year}{2017};\bibinfo{volume}{96}(\bibinfo{number}{3}):\bibinfo{pages}{032205}.
\bibitem[{Qiu et~al.(2019{\natexlab{b}})Qiu, Sun, Rudas and
  Gao}]{qiu2019command}
\bibinfo{author}{Qiu\xfnm[ J.]}, \bibinfo{author}{Sun\xfnm[ K.]},
  \bibinfo{author}{Rudas\xfnm[ I.J.]}, \bibinfo{author}{Gao\xfnm[ H.]}.
\newblock \bibinfo{title}{Command filter-based adaptive nn control for mimo
  nonlinear systems with full-state constraints and actuator hysteresis}.
\newblock \bibinfo{journal}{IEEE transactions on cybernetics}
  \bibinfo{year}{2019}{\natexlab{b}};.
\bibitem[{Pyragas and Pyragas(2019)}]{pyragas2019state}
\bibinfo{author}{Pyragas\xfnm[ V.]}, \bibinfo{author}{Pyragas\xfnm[ K.]}.
\newblock \bibinfo{title}{State-dependent act-and-wait time-delayed feedback
  control algorithm}.
\newblock \bibinfo{journal}{Communications in Nonlinear Science and Numerical
  Simulation}
  \bibinfo{year}{2019};\bibinfo{volume}{73}:\bibinfo{pages}{338--350}.
\bibitem[{Mainzer and Chua(2012)}]{Mainzer12}
\bibinfo{author}{Mainzer\xfnm[ K.]}, \bibinfo{author}{Chua\xfnm[ L.]}.
\newblock \bibinfo{title}{The Universe as Automaton: From Simplicity and
  Symmetry to Complexity}.
\newblock \bibinfo{publisher}{Springer}; \bibinfo{year}{2012}.
\bibitem[{Parker and Chua(2012)}]{parker2012practical}
\bibinfo{author}{Parker\xfnm[ T.S.]}, \bibinfo{author}{Chua\xfnm[ L.]}.
\newblock \bibinfo{title}{Practical numerical algorithms for chaotic systems}.
\newblock \bibinfo{publisher}{Springer Science \& Business Media};
  \bibinfo{year}{2012}.
\bibitem[{Benettin et~al.(1980)Benettin, Galgani, Giorgilli and
  Strelcyn}]{benettin1980lyapunov}
\bibinfo{author}{Benettin\xfnm[ G.]}, \bibinfo{author}{Galgani\xfnm[ L.]},
  \bibinfo{author}{Giorgilli\xfnm[ A.]}, \bibinfo{author}{Strelcyn\xfnm[
  J.M.]}.
\newblock \bibinfo{title}{{L}yapunov characteristic exponents for smooth
  dynamical systems and for hamiltonian systems; a method for computing all of
  them. part 1: theory}.
\newblock \bibinfo{journal}{Meccanica}
  \bibinfo{year}{1980};\bibinfo{volume}{15}(\bibinfo{number}{1}):\bibinfo{pages}{9--20}.
\bibitem[{Wolf et~al.(1985)Wolf, Swift, Swinney and
  Vastano}]{wolf1985determining}
\bibinfo{author}{Wolf\xfnm[ A.]}, \bibinfo{author}{Swift\xfnm[ J.B.]},
  \bibinfo{author}{Swinney\xfnm[ H.L.]}, \bibinfo{author}{Vastano\xfnm[ J.A.]}.
\newblock \bibinfo{title}{Determining {L}yapunov exponents from a time series}.
\newblock \bibinfo{journal}{Physica D: Nonlinear Phenomena}
  \bibinfo{year}{1985};\bibinfo{volume}{16}(\bibinfo{number}{3}):\bibinfo{pages}{285--317}.
\bibitem[{Dieci et~al.(1997)Dieci, Russell and Van~Vleck}]{dieci1997compuation}
\bibinfo{author}{Dieci\xfnm[ L.]}, \bibinfo{author}{Russell\xfnm[ R.D.]},
  \bibinfo{author}{Van~Vleck\xfnm[ E.S.]}.
\newblock \bibinfo{title}{On the compuation of {L}yapunov exponents for
  continuous dynamical systems}.
\newblock \bibinfo{journal}{SIAM Journal on Numerical Analysis}
  \bibinfo{year}{1997};\bibinfo{volume}{34}(\bibinfo{number}{1}):\bibinfo{pages}{402--423}.
\bibitem[{Stefanski(2000)}]{stefanski2000estimation}
\bibinfo{author}{Stefanski\xfnm[ A.]}.
\newblock \bibinfo{title}{Estimation of the largest {L}yapunov exponent in
  systems with impacts}.
\newblock \bibinfo{journal}{Chaos, Solitons \& Fractals}
  \bibinfo{year}{2000};\bibinfo{volume}{11}(\bibinfo{number}{15}):\bibinfo{pages}{2443--2451}.
\bibitem[{M{\"u}ller(1995)}]{muller1995calculation}
\bibinfo{author}{M{\"u}ller\xfnm[ P.C.]}.
\newblock \bibinfo{title}{Calculation of {L}yapunov exponents for dynamic
  systems with discontinuities}.
\newblock \bibinfo{journal}{Chaos, Solitons \& Fractals}
  \bibinfo{year}{1995};\bibinfo{volume}{5}(\bibinfo{number}{9}):\bibinfo{pages}{1671--1681}.
\bibitem[{Dellago et~al.(1996)Dellago, Posch and Hoover}]{dellago1996lyapunov}
\bibinfo{author}{Dellago\xfnm[ C.]}, \bibinfo{author}{Posch\xfnm[ H.A.]},
  \bibinfo{author}{Hoover\xfnm[ W.G.]}.
\newblock \bibinfo{title}{{L}yapunov instability in a system of hard disks in
  equilibrium and nonequilibrium steady states}.
\newblock \bibinfo{journal}{Physical Review E}
  \bibinfo{year}{1996};\bibinfo{volume}{53}(\bibinfo{number}{2}):\bibinfo{pages}{1485}.
\bibitem[{Jin et~al.(2006)Jin, Lu and Twizell}]{jin2006method}
\bibinfo{author}{Jin\xfnm[ L.]}, \bibinfo{author}{Lu\xfnm[ Q.]},
  \bibinfo{author}{Twizell\xfnm[ E.]}.
\newblock \bibinfo{title}{A method for calculating the spectrum of {L}yapunov
  exponents by local maps in non-smooth impact-vibrating systems}.
\newblock \bibinfo{journal}{Journal of Sound and Vibration}
  \bibinfo{year}{2006};\bibinfo{volume}{298}(\bibinfo{number}{4-5}):\bibinfo{pages}{1019--1033}.
\bibitem[{Lamba and Budd(1994)}]{lamba1994scaling}
\bibinfo{author}{Lamba\xfnm[ H.]}, \bibinfo{author}{Budd\xfnm[ C.]}.
\newblock \bibinfo{title}{Scaling of {L}yapunov exponents at nonsmooth
  bifurcations}.
\newblock \bibinfo{journal}{Physical Review E}
  \bibinfo{year}{1994};\bibinfo{volume}{50}(\bibinfo{number}{1}):\bibinfo{pages}{84}.
\bibitem[{Farmer(1982)}]{farmer1982chaotic}
\bibinfo{author}{Farmer\xfnm[ J.D.]}.
\newblock \bibinfo{title}{Chaotic attractors of an infinite-dimensional
  dynamical system}.
\newblock \bibinfo{journal}{Physica D: Nonlinear Phenomena}
  \bibinfo{year}{1982};\bibinfo{volume}{4}(\bibinfo{number}{3}):\bibinfo{pages}{366--393}.
\bibitem[{P{\'a}ez~Ch{\'a}vez et~al.(2020)P{\'a}ez~Ch{\'a}vez, Zhang and
  Liu}]{chavez2020numerical}
\bibinfo{author}{P{\'a}ez~Ch{\'a}vez\xfnm[ J.]}, \bibinfo{author}{Zhang\xfnm[
  Z.]}, \bibinfo{author}{Liu\xfnm[ Y.]}.
\newblock \bibinfo{title}{A numerical approach for the bifurcation analysis of
  nonsmooth delay equations}.
\newblock \bibinfo{journal}{Communications in Nonlinear Science and Numerical
  Simulation}
  \bibinfo{year}{2020};\bibinfo{volume}{83}:\bibinfo{pages}{105095}.
\bibitem[{Repin(1965)}]{repin1965approximate}
\bibinfo{author}{Repin\xfnm[ I.M.]}.
\newblock \bibinfo{title}{On the approximate replacement of systems with lag by
  ordinary dynamical systems}.
\newblock \bibinfo{journal}{Journal of Applied Mathematics and Mechanics}
  \bibinfo{year}{1965};\bibinfo{volume}{29}(\bibinfo{number}{2}):\bibinfo{pages}{254--264}.
\bibitem[{Gy{\"o}ri and Turi(1991)}]{gyori1991uniform}
\bibinfo{author}{Gy{\"o}ri\xfnm[ I.]}, \bibinfo{author}{Turi\xfnm[ J.]}.
\newblock \bibinfo{title}{Uniform approximation of a nonlinear delay equation
  on infinite intervals}.
\newblock \bibinfo{journal}{Nonlinear Analysis: Theory, Methods \&
  Applications}
  \bibinfo{year}{1991};\bibinfo{volume}{17}(\bibinfo{number}{1}):\bibinfo{pages}{21--29}.
\bibitem[{Breda(2006)}]{breda2006solution}
\bibinfo{author}{Breda\xfnm[ D.]}.
\newblock \bibinfo{title}{Solution operator approximations for characteristic
  roots of delay differential equations}.
\newblock \bibinfo{journal}{Applied Numerical Mathematics}
  \bibinfo{year}{2006};\bibinfo{volume}{56}(\bibinfo{number}{3-4}):\bibinfo{pages}{305--317}.
\bibitem[{Breda et~al.(2005)Breda, Maset and
  Vermiglio}]{breda2005pseudospectral}
\bibinfo{author}{Breda\xfnm[ D.]}, \bibinfo{author}{Maset\xfnm[ S.]},
  \bibinfo{author}{Vermiglio\xfnm[ R.]}.
\newblock \bibinfo{title}{Pseudospectral differencing methods for
  characteristic roots of delay differential equations}.
\newblock \bibinfo{journal}{SIAM Journal on Scientific Computing}
  \bibinfo{year}{2005};\bibinfo{volume}{27}(\bibinfo{number}{2}):\bibinfo{pages}{482--495}.
\bibitem[{Breda et~al.(2012)Breda, Maset and
  Vermiglio}]{breda2012approximation}
\bibinfo{author}{Breda\xfnm[ D.]}, \bibinfo{author}{Maset\xfnm[ S.]},
  \bibinfo{author}{Vermiglio\xfnm[ R.]}.
\newblock \bibinfo{title}{Approximation of eigenvalues of evolution operators
  for linear retarded functional differential equations}.
\newblock \bibinfo{journal}{SIAM Journal on Numerical Analysis}
  \bibinfo{year}{2012};\bibinfo{volume}{50}(\bibinfo{number}{3}):\bibinfo{pages}{1456--1483}.
\bibitem[{Chatelin(2011)}]{chatelin2011spectral}
\bibinfo{author}{Chatelin\xfnm[ F.]}.
\newblock \bibinfo{title}{Spectral approximation of linear operators}.
\newblock \bibinfo{publisher}{SIAM}; \bibinfo{year}{2011}.
\bibitem[{de~Souza et~al.(2008)de~Souza, Caldas, Viana and Balthazar}]{souza08}
\bibinfo{author}{de~Souza\xfnm[ S.L.T.]}, \bibinfo{author}{Caldas\xfnm[ I.L.]},
  \bibinfo{author}{Viana\xfnm[ R.L.]}, \bibinfo{author}{Balthazar\xfnm[ J.M.]}.
\newblock \bibinfo{title}{Control and chaos for vibro-impact and non-ideal
  oscillators}.
\newblock \bibinfo{journal}{J Theor Appl Mech}
  \bibinfo{year}{2008};\bibinfo{volume}{46}:\bibinfo{pages}{641--664}.
\bibitem[{Lazarek et~al.(2020)Lazarek, Brzeski, Solecki and
  Perlikowski}]{lazarek20}
\bibinfo{author}{Lazarek\xfnm[ M.]}, \bibinfo{author}{Brzeski\xfnm[ P.]},
  \bibinfo{author}{Solecki\xfnm[ W.]}, \bibinfo{author}{Perlikowski\xfnm[ P.]}.
\newblock \bibinfo{title}{Detection and classification of solutions for systems
  interacting by soft impacts with sample-based method}.
\newblock \bibinfo{journal}{Int J Bifur Chaos}
  \bibinfo{year}{2020};\bibinfo{volume}{30}:\bibinfo{pages}{2050079}.
\bibitem[{Serdukova et~al.(2020)Serdukova, Kuske and
  Yurchenko}]{serdukova2020postgrazing}
\bibinfo{author}{Serdukova\xfnm[ L.]}, \bibinfo{author}{Kuske\xfnm[ R.]},
  \bibinfo{author}{Yurchenko\xfnm[ D.]}.
\newblock \bibinfo{title}{Post-grazing dynamics of a vibro-impacting energy
  generator} \bibinfo{year}{2020};\bibinfo{volume}{arXiv.2003.02167}.
\bibitem[{Makarenkov and Lamb(2012)}]{makarenkov12}
\bibinfo{author}{Makarenkov\xfnm[ O.]}, \bibinfo{author}{Lamb\xfnm[ J.S.W.]}.
\newblock \bibinfo{title}{Dynamics and bifurcations of nonsmooth systems: {A}
  survey}.
\newblock \bibinfo{journal}{Physica D}
  \bibinfo{year}{2012};\bibinfo{volume}{241}:\bibinfo{pages}{1826--1844}.
\bibitem[{Liu and P{\'a}ez~Ch{\'a}vez(2017)}]{liu2017controlling}
\bibinfo{author}{Liu\xfnm[ Y.]}, \bibinfo{author}{P{\'a}ez~Ch{\'a}vez\xfnm[
  J.]}.
\newblock \bibinfo{title}{Controlling coexisting attractors of an impacting
  system via linear augmentation}.
\newblock \bibinfo{journal}{Physica D: Nonlinear Phenomena}
  \bibinfo{year}{2017};\bibinfo{volume}{348}:\bibinfo{pages}{1--11}.
\bibitem[{Liu et~al.(2013{\natexlab{a}})Liu, Wiercigroch, Pavlovskaia and
  Yu}]{liu2013modelling}
\bibinfo{author}{Liu\xfnm[ Y.]}, \bibinfo{author}{Wiercigroch\xfnm[ M.]},
  \bibinfo{author}{Pavlovskaia\xfnm[ E.]}, \bibinfo{author}{Yu\xfnm[ H.]}.
\newblock \bibinfo{title}{Modelling of a vibro-impact capsule system}.
\newblock \bibinfo{journal}{Int J Mech Sci}
  \bibinfo{year}{2013}{\natexlab{a}};\bibinfo{volume}{66}:\bibinfo{pages}{2--11}.
\bibitem[{P\'{a}ez~Ch\'{a}vez et~al.(2016)P\'{a}ez~Ch\'{a}vez, Liu, Pavlovskaia
  and M.}]{Chavez16}
\bibinfo{author}{P\'{a}ez~Ch\'{a}vez\xfnm[ J.]}, \bibinfo{author}{Liu\xfnm[
  Y.]}, \bibinfo{author}{Pavlovskaia\xfnm[ E.]}, \bibinfo{author}{M.\xfnm[
  W.]}.
\newblock \bibinfo{title}{Path-following analysis of the dynamical response of
  a piecewise-linear capsule system}.
\newblock \bibinfo{journal}{Comm Nonlinear Sci}
  \bibinfo{year}{2016};\bibinfo{volume}{37}:\bibinfo{pages}{102--114}.
\bibitem[{Liu et~al.(2013{\natexlab{b}})Liu, Pavlovskaia, Hendry and
  Wiercigroch}]{liu2013friction}
\bibinfo{author}{Liu\xfnm[ Y.]}, \bibinfo{author}{Pavlovskaia\xfnm[ E.]},
  \bibinfo{author}{Hendry\xfnm[ D.]}, \bibinfo{author}{Wiercigroch\xfnm[ M.]}.
\newblock \bibinfo{title}{Vibro-impact responses of capsule system with various
  friction models}.
\newblock \bibinfo{journal}{Int J Mech Sci}
  \bibinfo{year}{2013}{\natexlab{b}};\bibinfo{volume}{72}:\bibinfo{pages}{39--54}.
\bibitem[{Liu and P\'{a}ez~Ch\'{a}vez(2017)}]{Liu17}
\bibinfo{author}{Liu\xfnm[ Y.]}, \bibinfo{author}{P\'{a}ez~Ch\'{a}vez\xfnm[
  J.]}.
\newblock \bibinfo{title}{Controlling multistability in a vibro-impact capsule
  system}.
\newblock \bibinfo{journal}{Nonlinear Dyn}
  \bibinfo{year}{2017};\bibinfo{volume}{88}:\bibinfo{pages}{1289--1304}.
\bibitem[{Wojewoda et~al.(2008)Wojewoda, Stefański, Wiercigroch and
  Kapitaniak}]{Wojewoda08}
\bibinfo{author}{Wojewoda\xfnm[ J.]}, \bibinfo{author}{Stefański\xfnm[ A.]},
  \bibinfo{author}{Wiercigroch\xfnm[ M.]}, \bibinfo{author}{Kapitaniak\xfnm[
  T.]}.
\newblock \bibinfo{title}{Hysteretic effects of dry friction: modelling and
  experimental studies}.
\newblock \bibinfo{journal}{Philosophical Transactions of the Royal Society A:
  Mathematical, Physical and Engineering Sciences}
  \bibinfo{year}{2008};\bibinfo{volume}{366}:\bibinfo{pages}{747--765}.
\bibitem[{Pyragas(1992)}]{pyragas1992continuous}
\bibinfo{author}{Pyragas\xfnm[ K.]}.
\newblock \bibinfo{title}{Continuous control of chaos by self-controlling
  feedback}.
\newblock \bibinfo{journal}{Physics Letters A}
  \bibinfo{year}{1992};\bibinfo{volume}{170}(\bibinfo{number}{6}):\bibinfo{pages}{421--428}.
\bibitem[{Guo et~al.(2020)Guo, Liu, Birler and Prasad}]{Guo20}
\bibinfo{author}{Guo\xfnm[ B.]}, \bibinfo{author}{Liu\xfnm[ Y.]},
  \bibinfo{author}{Birler\xfnm[ R.]}, \bibinfo{author}{Prasad\xfnm[ S.]}.
\newblock \bibinfo{title}{Self-propelled capsule endoscopy for small-bowel
  examination: proof-of-concept and model verification}.
\newblock \bibinfo{journal}{Int J Mech Sci}
  \bibinfo{year}{2020};\bibinfo{volume}{174}:\bibinfo{pages}{105506}.
\bibitem[{Krasovskii(1962)}]{krasovskii1962analytic}
\bibinfo{author}{Krasovskii\xfnm[ N.]}.
\newblock \bibinfo{title}{On the analytic construction of an optimal control in
  a system with time lags}.
\newblock \bibinfo{journal}{Journal of Applied Mathematics and Mechanics}
  \bibinfo{year}{1962};\bibinfo{volume}{26}(\bibinfo{number}{1}):\bibinfo{pages}{50--67}.
\bibitem[{Banks(1979)}]{banks1979approximation}
\bibinfo{author}{Banks\xfnm[ H.]}.
\newblock \bibinfo{title}{Approximation of nonlinear functional differential
  equation control systems}.
\newblock \bibinfo{journal}{Journal of Optimization Theory and Applications}
  \bibinfo{year}{1979};\bibinfo{volume}{29}(\bibinfo{number}{3}):\bibinfo{pages}{383--408}.
\bibitem[{Stoer and Bulirsch(2013)}]{stoer2013introduction}
\bibinfo{author}{Stoer\xfnm[ J.]}, \bibinfo{author}{Bulirsch\xfnm[ R.]}.
\newblock \bibinfo{title}{Introduction to numerical analysis}.
\newblock \bibinfo{publisher}{Springer Science \& Business Media};
  \bibinfo{year}{2013}.
\bibitem[{Jiang and Wiercigroch(2016)}]{jiang2016geometrical}
\bibinfo{author}{Jiang\xfnm[ H.]}, \bibinfo{author}{Wiercigroch\xfnm[ M.]}.
\newblock \bibinfo{title}{Geometrical insight into non-smooth bifurcations of a
  soft impact oscillator}.
\newblock \bibinfo{journal}{IMA Journal of Applied Mathematics}
  \bibinfo{year}{2016};\bibinfo{volume}{81}(\bibinfo{number}{4}):\bibinfo{pages}{662--678}.
\bibitem[{Chatelin(1973)}]{chatelin1973convergence}
\bibinfo{author}{Chatelin\xfnm[ F.]}.
\newblock \bibinfo{title}{Convergence of approximation methods to compute
  eigenelements of linear operations}.
\newblock \bibinfo{journal}{SIAM Journal on Numerical Analysis}
  \bibinfo{year}{1973};\bibinfo{volume}{10}(\bibinfo{number}{5}):\bibinfo{pages}{939--948}.
\bibitem[{Varma and Mills(1973)}]{varma1973summability}
\bibinfo{author}{Varma\xfnm[ A.]}, \bibinfo{author}{Mills\xfnm[ T.]}.
\newblock \bibinfo{title}{On the summability of lagrange interpolation}.
\newblock \bibinfo{journal}{Journal of Approximation Theory}
  \bibinfo{year}{1973};\bibinfo{volume}{9}(\bibinfo{number}{4}):\bibinfo{pages}{349--356}.

\end{thebibliography}

\end{document}